\documentclass{aa}   
\usepackage{graphicx} 
\usepackage{natbib} 
\usepackage{txfonts} 
\usepackage{multirow} 
\begin{document} 
   \title{A multi-wavelength view of AB Dor's outer atmosphere \thanks{Based on observations 
collected at the European Southern Observatory, Paranal, Chile, 383.D-1002A and on observations 
obtained with \emph{XMM-Newton}, an ESA science mission with instruments and contributions directly funded  
by ESA Member states and NASA.} 
   \subtitle{Simultaneous X-ray and optical spectroscopy at high cadence}} 
 
   \author{S. Lalitha$^{1}$, B. Fuhrmeister$^{1}$, U. Wolter$^{1}$, J. H. M. M. Schmitt$^{1}$, D. Engels$^{1}$, \and M. H. Wieringa$^{2}$ 
          } 
   \authorrunning{S. Lalitha et al.} 
   \titlerunning{AB Doradus - outer atmospheres} 
   \offprints{lsairam@hs.uni-hamburg.de} 
 
   \institute{$^1$Hamburger Sternwarte, University of Hamburg, 
              Gojenbergsweg 112, 21029 Hamburg, Germany\\ 
	      $^2$CSIRO Astronomy \& Space Science, Locked Bag 194,Narrabri, NSW 2390, Australia\\ 
              \email{lsairam@hs.uni-hamburg.de} 
               } 
 
   \date{Received XXXX; accepted XXXX}

\abstract    
 {} 
{ We study the chromosphere and corona of the ultra-fast rotator 
  \object{AB~Dor~A} at high temporal and spectral resolution using simultaneous 
  observations with \emph{XMM-Newton} in the X-rays, VLT/UVES in the optical, and the ATCA in 
  the radio. Our optical 
  spectra have a resolving power of $\sim$ 50\,000 with a time cadence of 
  $\sim$ 1 min. 
  Our observations continuously cover more than one rotational period   
  and include both quiescent periods and three flaring events of different strengths.} 
{From the X-ray observations we investigated the variations in coronal temperature, emission measure, 
densities, and abundance. We interpreted our data in terms of a loop model.  
   From the optical data we characterise the flaring chromospheric material using numerous emission lines that appear in the course of the flares. A detailed analysis of the 
line shapes and line centres allowed us to infer physical characteristics of the flaring chromosphere 
and to coarsely localise the flare event on the star.} 
   {We specifically used the optical high-cadence spectra to demonstrate that both, turbulent 
and Stark broadening are present during the first ten minutes of the first flare. Also, 
in the first few minutes of this flare, we find short-lived (one to several minutes) 
emission subcomponents in the H$\alpha$ and \ion{Ca}{ii} K lines, which we interpret as flare-connected shocks owing to their high intrinsic velocities. 
Combining the space-based data with the results of our optical spectroscopy, 
we derive flare-filling factors. Finally, comparing X-ray, optical broadband, and line emission, 
we find a correlation for two of the three flaring events, while there is no clear correlation for one event. Also, we do not find any 
correlation of the radio data 
to any other observed data.} 
{} 
 
 
   \maketitle 
%

\section{Introduction}

Low mass stars possess stratified atmospheres with coronae, transition regions, 
chromospheres and photospheres with characteristic temperatures and signatures of
magnetically induced activity. Photospheric spots have historically provided 
the first evidence of (magnetic) activity on the Sun and the sunspot records 
were later complemented by observations in other spectral bands at X-ray, UV, and radio
wavelengths, which trace different layers of the atmosphere, hence different activity phenomena. 
These different layers of the Sun's atmosphere are not physically independent, for instance during flares 
chromospheric material is mixed into the corona, causing a 
temporary change in coronal metallicity (\citealt{sylwester}, also see \citealt{phillips} and Sect. 6 of \citealt{fletcher}).  
Moreover, for the case of the \emph{spatially resolved} quiescent Sun,  
\citet{beck}  found propagating events between the photosphere and the 
chromosphere with travel times of about 100 s. 
The different layers are connected by magnetic flux tubes, which arc into the corona as loops \citep{fossum, Wedemeyer}. 
  
For the Sun, the heating mechanisms of its outer layers are not understood well \citep[e.g.][]{Tornados}. 
For other stars, our understanding is even coarser because their activity phenomena cannot be spatially resolved in most cases. 
Furthermore, stellar observations often lack  either the temporal or the spectral 
resolution required to follow the fast changing and complex phenomena in active regions. 
  
To our knowledge our study is the first that combines high resolution 
(R $\sim$ 50\,000) and high-cadence ($\sim$ 50 s with 25 s exposure time)  optical 
spectra with simultaneous X-ray data that also offer at least this temporal resolution.  
Other flare studies usually concentrate on high resolution 
\textit{or} high cadence; e.\,g. \citet{Crespo} studied flares on AD~Leo with 
a cadence of typically 3 min, but with R of $\sim$ 3500 in the wavelength 
range of 3500 to 5176 \AA. 
Also \citet{Kowalski} observed a white light mega-flare on YZ~CMi with a 
resolving power below 1000 but a cadence of 30 seconds covering a 
wavelength range from 3350 to 9260 \AA. 
Prominent examples of good spectral 
resolution but lower cadence are \citet{Montes}, who observed a major flare on LQ~Hya 
during decay and about half an hour before its onset 
with R $\sim$ 35\,000 and a cadence of six to seven minutes in the wavelength range between 
4842 to 7725 \AA. \citet{CNLeoflare} observed CN~Leo during a 
mega flare with a cadence of about 15 minutes and R$\sim$ 50\,000. 
  
Achieving both high spectral resolution and high cadence as in our study is only possible 
through the combination of the instruments used and a bright target, 
AB~Dor~A, an extremely active, young K-dwarf and the closest ultra-fast rotator 
($P_{rot}$ =  0.51 days, d = 14.9\,pc, $m_v$ = 6.9\,mag, see \citealt{guirado} and 
their references).  

AB~Dor~A is a member of a quadruple system. The visual companion of  
AB~Dor~A is an active M-dwarf -- Rst~137B or AB~Dor~B 
\citep{vilhu1}, located $\sim$9.5$\arcsec$ away from AB~Dor~A and
$\sim$60 times bolometrically fainter than the primary.  
The binarity of AB~Dor~B (with a separation of $\sim$0.7$\arcsec$) was only discovered 
after the advent of the adaptive optics \citep{close_2005}. 
The third component, AB~Dor~C, is a low mass companion
\citep{close_2007} located $\sim0.16\arcsec$ away from AB~Dor~A. 
The contribution from the companions to the spectrum of AB~Dor~A can be assumed to be 
negligible because of their relative faintness.

AB~Dor~A is well studied target 
across all wavelengths which demonstrates its high activity level.  
At longer wavelengths, AB~Dor~A was found to be a highly variable radio source by \citet{white}.  
It also showed strong evidence of rotational modulation in its
radio emission \citep{lim}.  
At optical wavelengths, signs of photospheric activity were found in the form of
long-lived spots \citep{pakull, innis1} and by magnetic fields with a typical field strengths of 
$\approx$500\,G covering about 20$\%$ of the surface \citep{donati}. 
At X-ray wavelengths, AB~Dor~A has been observed frequently ever
since its detection by the {\it Einstein Observatory} \citep{pakull,vilhu1}. 
Later observations were carried out with ROSAT \citep{Kuerster}, \emph{XMM-Newton} \citep{guedel, sanz}, and \emph{Chandra} \citep{sanz, hussain, garcia}.  
The corona of AB~Dor~A shows a high level of variability with frequent flaring on time scales from 
minutes to hours. \citet{vilhu} estimated  an average of at least one flare per rotation on 
AB~Dor~A's surface.

Our paper is structured as follows. In Sect.~2 we describe our observations obtained in the
three wavelength bands, and
in Sect.~3 we compare the temporal behaviour of AB~Dor~A at radio, optical, and soft X-ray wavelengths. 
We present  the coronal properties of AB~Dor~A in quiescence, as well as during flaring state in Sect.~4, 
while in Sect.~5 we describe the chromospheric properties of the star. Sections~6 and 7 contain the 
discussion and our conclusions.


\section{Observations and data analysis} 
 
The data on AB~Dor~A discussed in this paper  
were obtained simultaneously with \emph{XMM-Newton}, ESO's Kueyen  
telescope equipped with the Ultraviolet-Visual Echelle Spectrograph 
(UVES) and the Australian Compact Array (ATCA)\footnote{The Australia Telescope Compact Array is part of the Australia Telescope 
National Facility, which is funded by the Commonwealth of Australia for 
operation as a National Facility managed by CSIRO.} on 25/26 November 2009. 
  
\subsection{Optical UVES data}\label{UVESdata}

For the optical data acquisition the UVES spectrograph was operated in a 
dichroic mode, leading to a spectral coverage from 3720~\AA\, to 4945~\AA\, in 
the blue arm and 5695~\AA\, to 9465~\AA\, in the red arm with a small gap from 
7532~\AA\, to 7655~\AA\, due to the CCD mosaic.\footnote{A detailed description of the UVES 
spectrograph is available at http://www.eso.org/instruments/uves/doc/} 
Data were taken on 26 November 2009 between 2:00 UT and 9:30 UT covering 
approximately 60\% of one rotational period. 
We used an exposure time of 25 s and achieved an effective time 
resolution of about 
50 s owing to the CCD readout, resulting in 460 spectra for the whole night.  
The resolution of our spectra is $\sim$ 40\,000 for the blue spectra 
and $\sim$ 60\,000 for the red spectra. The signal-to-noise ratio (S/N) 
for the blue spectra is about 100 and for the red spectra 150-200.
There are two short data gaps, one at about 6:55 UT due to technical problems, 
when two spectra were lost and one at about 8:00 UT due to observations of AB~Dor~B. 
The spectra were reduced using the UVES pipeline vers. 4.4.8 \footnote{The~UVES pipeline~manual~can be found at  
ftp://ftp.eso.org/pub/dfs/pipelines/uves/uves-pipeline-manual-13.0.pdf} 
including wavelength and flux calibration. 
 
We removed the telluric lines around H$\alpha$ line using a table 
of telluric water lines \citep{Clough}. We broadened the lines with a Gaussian representing 
the instrumental resolution and used a telluric reference line at 6552.61~\AA~  for 
fitting the FWHM of the instrumental Gaussian and the depth of the line to 
each of our spectra, and then subtracted the telluric spectrum.

\subsection{X-ray data} 
 
The X-ray data were obtained using the \emph{XMM-Newton} \footnote{A detailed~description~of the instruments on-board \emph{XMM-Newton} is available at  
http://xmm.esac.esa.int/external/xmm\_user\_support/documentation/technical/.}  
Observatory. 
\emph{XMM-Newton} carries three co-aligned X-ray telescopes. 
Each of the three telescopes is equipped with a CCD camera, and together then form the European Photon Imaging Camera  
(EPIC). One of the telescopes is equipped with a pn CCD, and the other two telescopes  
carry an MOS (Metal Oxide Semi-conductor) CCD each with sensitivity between $\sim$ 0.2 keV and 15 keV. 
These X-ray CCD detectors provide medium resolution imaging spectroscopy  
(E/$\delta$E $\sim$ 20-50) and a temporal resolution at subsecond level. 
The telescopes with the MOS detectors are  
equipped with reflection gratings that provide simultaneous high resolution X-ray spectra 
between 0.35 and 2.5~keV with the Reflection Grating Spectrometer (RGS). 
In addition, \emph{XMM-Newton} carries an Optical Monitor (OM), which is an optical/UV telescope 
with different filters  for imaging and time-resolved photometry.  
  
Our XMM-Newton observations have a total duration 
of $\sim$58~ks, with data being taken between 21:00 UT on 25 November 2009 and 13:06 UT on 26 November 2009 (Obs ID: 
0602240201), covering 1.3 times the rotational period and, in particular, the 
entire time span of our optical observations. 
Useful data of AB~Dor~A were obtained with the OM, EPIC, and the 
RGS detectors, which were all operated simultaneously. 
The pn and MOS detectors were operated with the medium filter in 
imaging and small window mode.  
The OM was operated in fast mode with 0.5 s cadence using the UVM2 band filter 
covering a band pass between 2050 - 2450 \ \AA. 
 
All X-ray data were reduced with the \emph{XMM-Newton} Science Analysis System (SAS) 
\footnote{The \emph{XMM-Newton} SAS user guide can be found at  
http://xmm.esac.esa.int/external/xmm\_user\_support/documentation/} 
 software, version 
12.0.1. 
EPIC light curves and spectra were obtained using standard 
filtering criteria. Spectral analysis was performed using XSPEC 
version 12.5.0 \citep{xspec} for the overall fitting processes.  
The models we used for fitting assume a collisionally ionised optically thin  
plasma as calculated with the APEC code \citep{apec}, and elemental abundances are  
calculated relative to the solar photospheric values of \citet{Grevesse}. 
 
\subsection{Radio data} 
 
AB Dor was observed with the Australian Compact Array (ATCA) 
on 25 November 2009 from 19:00 UT until 26 November 2009 18:00 UT, which 
corresponds to 1.8 times the rotational period, 
with a major interruption between 00:08 and 06:18 UT on the 26$^{th}$ 
(for details on the instrument  
\citealt{wilson_2011}). 
The array was in configuration 6B with baselines up to 6000 meters, providing 
a spatial resolution of 1--2 arcsec for the observed frequencies. 
The back-end was centred on 5.5 and 9.0 GHz, and the bandwidth was 2 
GHz in both cases. Data was taken every 10 sec with breaks for 
calibrator (PKS 0515-674) scans every 7-15 minutes, depending on 
weather conditions.   
 
Data reduction was performed using the Miriad package \citep{sault}. 
 Time 
periods with bad phase stability and frequency channels affected by 
radio-frequency interference (RFI) were flagged. Bandpass calibration was performed using PKS 
0823-500 for the first half of the data and PKS 1934-638 for the 
second half. Phase and gain calibration was performed using the 
frequent observations of PKS 0515-674. Absolute amplitude calibration 
was performed using PKS 1934-638, assuming a flux density of 4.97 Jy 
at 5.5 GHz and 2.70 Jy at 9.0 GHz.

Images were generated at both frequencies using the full data set. The two 
strongest sources detected were at $\alpha$=5:28:44.95 
$\delta$=-65:26:53.5 and $\alpha$=5:28:44.61 $\delta$=-65:26:44.7 (J2000), 
which we identify as AB~Dor~A and B, respectively. The mean flux densities were 4.2 
mJy (5.5 GHz) and 3.0 mJy (9.0 GHz) for the primary star and 2.0 mJy 
(5.5 GHz) and 1.6 mJy (9.0 GHz) for AB~Dor~B. The noise level in the 
images was 11 $\mu$Jy and 12 $\mu$Jy at 5.5 and 9.0 GHz, respectively. 
 
Separate light curves in radio wavelength were produced for the two stars (AB~Dor~A and AB~Dor~B). They were phase-shifted 
to the field centre and then self-calibrated (three 
 iterations on the phases). After subtracting all other sources 
 in the field, the data was vector averaged over a time interval of 
 120 sec over all baselines and channels. 
 

\section{Temporal analysis}  
\label{sec:temp} 
 
\begin{figure*} 
\begin{center} 
\includegraphics[width=17.0cm,height=18.0cm,clip]{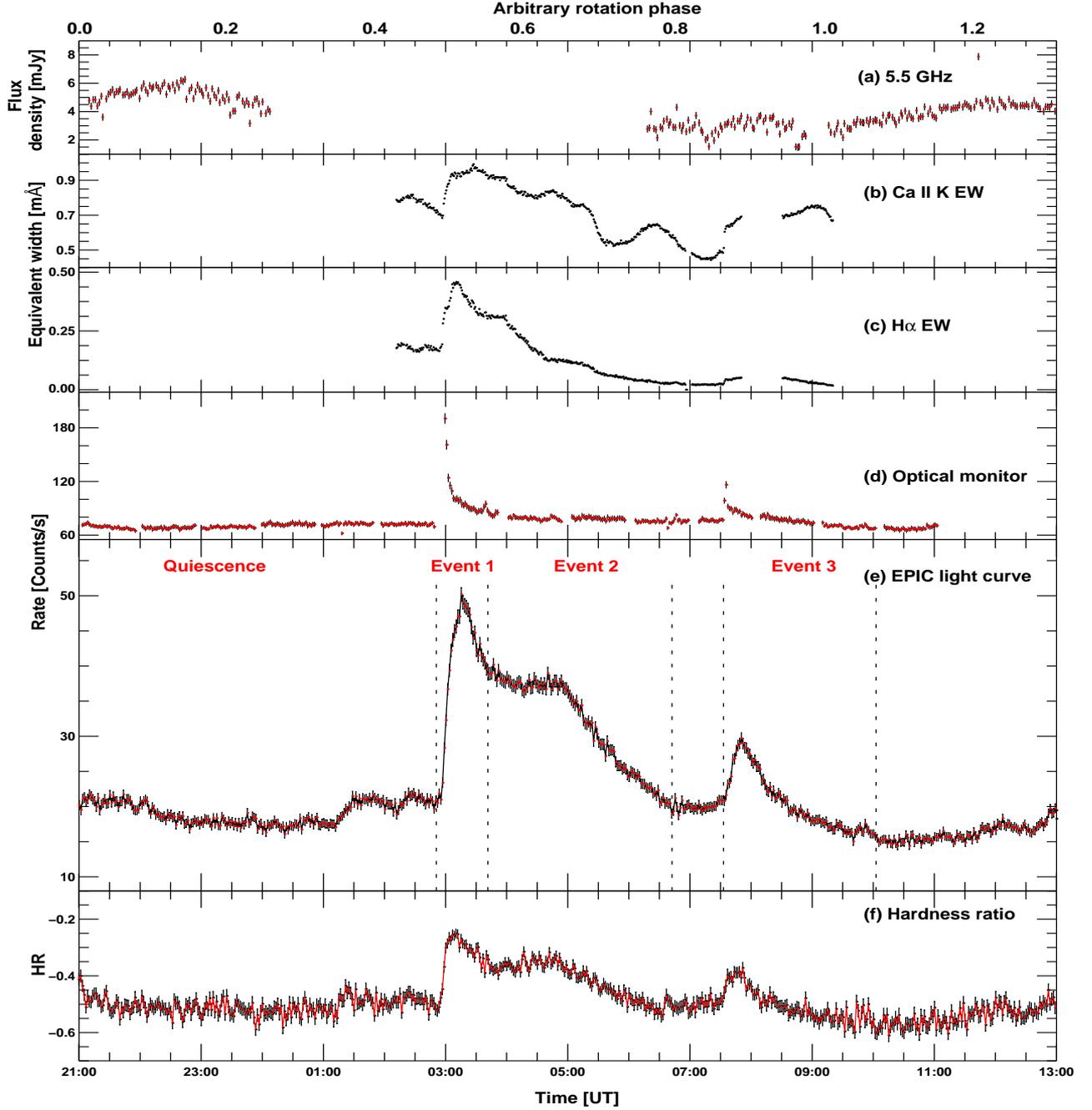}\vspace{-0.4mm} 
\caption{\label{light_curve} Light curves of AB~Dor~A observed on 25 Nov  
2009: (a) 5.5 GHz radio flux from ATCA observations binned to 120 s,  
(b) and (c) \ion{Ca}{ii} K flux and $H\alpha$ equivalent widths calculated 
 from UVES spectra, 
(d) OM light curve, (e) EPIC (combined MOS and pn) light curve, (f) the EPIC hardness ratio. Light curves (d) to (f) are binned to 100~s. 
The vertical lines in panel (e) indicate the time  
segment corresponding to the events discussed in the article. Plotted at the top is the arbitrary  
rotation phase.} 
\end{center} 
\end{figure*}

In Fig. \ref{light_curve} we provide a summary of our observations 
carried out as a multi-wavelength campaign designed to cover the coronal and chromospheric properties of AB~Dor~A.  
Starting from the top in Fig.~\ref{light_curve}, we plot the radio light curve recorded at 5.5 GHz with 120~s binning (panel a). 
The 9.0 GHz light curve is not shown, 
since it is very similar to the 5.5 GHz light curve.  
In Figs. \ref{light_curve}(b) and~\ref{light_curve}(c), we show the measured  
UVES \ion{Ca}{ii} K and H$\alpha$ equivalent widths (EW) as two examples of strong chromospheric emission lines, 
which originate in the lower and upper chromospheres, respectively.   
In Figs.~\ref{light_curve}(d) and~\ref{light_curve}(e), 
we plot the {\it XMM-Newton} OM and EPIC (pn and MOS combined) light curves with a binning of 100s.  
To identify heating events, we define a hardness ratio (HR) for the EPIC-pn as 
$HR = \small{\frac{H - S}{H + S}}$, where  H is the number of counts between 
1.0 and 10.0 keV (hard band), and S the number of counts between 0.15  
and 1.0 keV (soft band), and plot the time-dependent hardness ratios (HR) in Fig.~\ref{light_curve}(f). 
The most extensive data set comes from {\it XMM-Newton}, which observed AB~Dor for a total of 16 hours 
contiguously (from Nov 25 21:00 to Nov 26 13:00). UVES data are available for the time span between 
Nov 26 2:00 to 9:30, radio data is available from Nov 25 21:00-24:00~UT and Nov 26 6:00-13:00~UT.

The most notable feature in the AB~Dor light curve is a large flare or possibly a
sequence of flares lasting from about 
Nov 26 3:00-9:00~UT, which was covered by both {\it XMM-Newton} and UVES simultaneously, while the major part of the flare was unfortunately missed at radio wavelengths. A rough estimate of the flare 
energetics is given in Table \ref{energy}. There are in addition a number of small scale events  
visible in the  {\it XMM-Newton} OM and EPIC light curves as can be seen in 
Fig. \ref{light_curve}(e); however, we concentrate on the large events in this paper.  
\begin{table} 
\centering 
\caption{\label{energy} Integrated energies of individual flare events in XMM-Newton's EPIC and OM. } 
\begin{tabular}[htbp]{ccc} 
\hline 
\hline 
Instrument  & energy   [erg]  &    energy  [erg]\\ 
            & Event 1 + Event 2     & Event 3\\ 
            & 2:50-6:40 UT & 7:30-10:00 UT\\ 
\hline \\[-3mm] 
OM    & 4.20$\times10^{30}$ &1.40$\times10^{30}$\\ 
EPIC  & 3.57$\times10^{33}$ &9.40$\times10^{32}$\\ 
 
\hline 
\end{tabular} 
\end{table} 
 
For purposes of discussion we distinguish the following events, which may or may not be physically connected: 
 
\begin{itemize} 
 
\renewcommand{\labelitemi}{$\bullet$} 
\item Event 1: The first and main flare starts at 02:57 UT, marked by a steep flux rise in all covered wavelength bands. 
Also many chromospheric emission lines go into  
emission at this instance, e.\,g. all Balmer lines covered, the \ion{Ca}{ii} H and K line, the \ion{Na}{i} D lines, and the \ion{He}{i} D$_{3}$ line.  The X-ray count rate 
increased from a quiescent value of $\sim$ 14 cts/s in the pn and $\sim$ 5 cts/s in the MOS detector  
to $\sim$ 38 cts/s and $\sim$ 15 cts/s in the pn and MOS detectors, respectively, at flare peak.  
In the HR we find a clear hardening to $\approx -0.2$ during the large flare.  
The end of Event 1 cannot be defined uniquely since it is overlaid by Event 2. Therefore, we define its end simply as the onset of Event 2. 
 
\item Event 2: The second event has a broad, slowly changing light curve with  
its onset marked by a small flare-like event at about 
3:40 UT in the OM light curve, which coincides with a slope change in 
the chromospheric and X-ray light curves. The hardness ratio starts to increase 
again at about the same time. The decay of Event 2 lasts  
until  6:40 UT, when a short period of constant X-ray count rate starts.  
The pronounced broad maximum of Event 2 occurring in the X-ray 
data at about 5:00 UT has no obvious counterpart in the other wavelength bands. In the OM light curve there is 
again a minor event and the H$\alpha$ light curve displays a plateau at the time of 
the X-ray maximum 
(see Sect. \ref{X-ray-optical}). From the light curve morphology it is not clear whether Event 2 is associated  
with Event 1 as a reheating episode or whether it is independent so we pursue this issue 
further in Sect. \ref{discussion}. 
  
\item Event 3: The third event starts at 07:34 UT and lasts until about 10:00 UT.  
This event can be traced in the
X-ray and OM data and in the chromospheric emission lines. 
It is the only event with simultaneous coverage at radio wavelengths; however, 
the radio light curve shows no clear response when compared to the flare data 
in the other bands. 
 
\end{itemize} 
  
In addition to the observed flaring,  
the \ion{Ca}{ii} K EW curve between 5:00 UT and 8:00 UT  
shows two dimming events, which are usually interpreted to be caused by prominence crossings 
(see Sect. \ref{prominence}). 
The radio light curve does exhibit a dip in parallel to the second chromospheric dip. Since the first chromospheric dimming was not covered in radio 
data and the radio light curve shows signs of another dip at about 9:00 UT,
when no dimming is noted in other diagnostics, it is hard to decide whether 
indeed the radio signal is physically connected to the prominence or only a coincidence.  
In our H$\alpha$ EW light curve the information about the dimming events is lost   
due to the large line width chosen for determining the EW in order to cover 
the broad emission component emerging during Event 1. The EW light curve 
is totally dominated by this broad line component during Event 1; however, 
the dimming events can  nevertheless be seen in H$\alpha$, e.~\,g. in Fig. \ref{intensitymap}, where we provide  a more detailed discussion.

\section{X-ray spectral analysis} 
 
In the following section we provide a detailed discussion of our X-ray observations of AB~Dor. 
 
\subsection{Spectral fits and elemental abundances} 
\label{sec:corabund} 
\subsubsection{Quiescent and flaring emission} 
 
The time span before 2:50 UT is free of any large temporal variations, so we define this period as the preflare quiescent state of AB~Dor~A. 
The comparison of our EPIC-MOS spectra taken during quiescence and during the flare rise phase   
shows the expected flare-related changes in the spectral energy distribution,
yielding a substantial increase in emission at higher temperature. 
During a flare, fresh material from the photosphere and chromosphere is brought to the upper  
layers of the stellar atmosphere, 
resulting in a temporary change in the coronal abundance.  
 
We therefore performed a detailed X-ray spectral analysis to determine plasma temperatures,  
emission measures, and abundances for the different activity states of AB~Dor~A. 
We specifically determined the abundances  
relative to solar values \citep{Grevesse} 
with an iterative procedure of  
global XSPEC fits to EPIC and RGS spectra with VAPEC plasma models.  
In these fits, the derived abundances and emission measures are inherently interdependent, therefore we make our inferences from the relative changes of the fit parameters derived for the observations.  
 
We compare each of the events introduced in Sect.\ref{sec:temp} 
with the quiescent state using our RGS and EPIC-MOS spectra. 
We fit each of these spectra in the energy range 0.2-10.0 keV in the
following fashion. We first constructed a fixed temperature  
grid at the grid points $0.3$, $0.6$, $1.2$, and $2.4$~keV  
($\sim$3.5, $\sim$7.0, $\sim$14.0, and $\sim$28.0 MK,  
respectively).  
These temperature grids agree with the best fit temperatures obtained 
by \citet{sanz}, who analysed three years of XMM-Newton data of AB~Dor~A. 
For fitting RGS spectra, the abundance of carbon, nitrogen, oxygen,  
neon, and iron were allowed to vary independently, but were 
fixed between all VAPEC temperature components. 
Meanwhile when fitting EPIC-MOS spectra, the abundances of magnesium,  
silicon, sulphur, and argon were allowed to vary along with those of 
oxygen, neon, and iron, but the carbon and nitrogen abundances were 
fixed to values obtained from RGS, which is sensitive to strong individual lines;  
the resulting fit parameters for the various states of AB~Dor~A are given in Table \ref{obs}. 
Satisfactory fits (from a statistical point of view) can be obtained. During all the flares a pronounced enhancement of the emission measure at 
2.4 keV (28~MK) is present and, to a lesser extent, at the very softest energies.    
As far as the abundance pattern between flare and quiescent states is concerned,  
there is no clear difference except for the elements Fe and Ne,  
which are clearly enhanced  during the flaring states.

\begin{table} 
\begin{center} 
\caption{\label{obs} Fixed 4-temperature grid fit to the quiescent and the flaring states with variable elemental abundances.  
The abundances are relative to solar photospheric abundances \citep{Grevesse}.}   
\fontsize{8pt}{10pt}\selectfont  
\begin{tabular}[htbp]{llllll} 
\hline\hline 
Parameters & Quiescence & Event 1 & Event 2 & Event 3\\ 
\hline \\[-3mm]\\ 
$kT_1$ [keV] 	& \multicolumn{4}{c}{0.3}	\\ 
$EM_{1}$  	&$1.43 \pm 0.16$ & $2.12 \pm 0.32$   & $2.29 \pm 0.29$ & $2.37 \pm 0.16$\\	 
$[10^{52}$ $cm^{-3}]$ 	&&&&\\ 
$kT_2$ [keV]	& \multicolumn{4}{c}{0.6}	\\ 
$EM_{2}$ &$4.59 \pm 0.26$ & $4.35 \pm 0.46$	& $5.53 \pm 0.43$ & $4.34 \pm 0.24$\\	 
$[10^{52}$ $cm^{-3}]$ 	&&&&\\ 
$kT_3$ [keV] 	& \multicolumn{4}{c}{1.2}	\\ 
$EM_{3}$  	&$1.98 \pm 0.19$ & $1.16 \pm 0.44$	& $2.11 \pm 0.37$ & $1.72 \pm 0.16$\\ 
$[10^{52}$ $cm^{-3}]$ 	&&&&\\ 
$kT_4$ [keV]	& \multicolumn{4}{c}{2.4}	\\ 
$EM_{4}$  	&$1.79 \pm 0.11$ & $11.06 \pm 0.40$	& $8.47 \pm 0.29$ & $5.03 \pm 0.11$\\ 
$[10^{52}$ $cm^{-3}]$ 	&&&&\\ 
C 		& 0.71$ \pm 0.10$	&0.82$ \pm 0.20$ 	& 0.67 $\pm$ 0.13	& 0.69 $\pm$ 0.13\\ 
N 		& 0.90$ \pm 0.14$	&0.82$ \pm 0.24$	& 0.77 $\pm$ 0.19	& 0.96 $\pm$ 0.19 \\ 
O 		& 0.41$ \pm 0.02$	&0.38$ \pm 0.03$	& 0.34 $\pm$ 0.02	& 0.35 $\pm$ 0.01\\ 
Ne 		& 0.81$ \pm 0.04$	&1.02$ \pm 0.08$	& 0.93 $\pm$ 0.05	& 0.96 $\pm$ 0.04\\ 
Mg 		& 0.27$ \pm 0.03$	&0.22$ \pm 0.07$	& 0.28 $\pm$ 0.05	& 0.32 $\pm$ 0.03\\ 
Si 		& 0.22$ \pm 0.04$	&0.25$ \pm 0.08$	& 0.33 $\pm$ 0.05	& 0.27 $\pm$ 0.03\\ 
S 		& 0.31$ \pm 0.05$	&0.29$ \pm 0.09$	& 0.31 $\pm$ 0.07	& 0.33 $\pm$ 0.04\\ 
Ar 		& 1.02$ \pm 0.26$	&0.66$ \pm 0.40$	& 1.13 $\pm$ 0.32	& 0.76 $\pm$ 0.20\\ 
Fe 		& 0.25$ \pm 0.01$	&0.32$ \pm 0.02$	& 0.26 $\pm$ 0.01	& 0.29 $\pm$ 0.01\\ 
red. $\chi^{2}$	&1.13				&1.09		& 1.12			&1.68 \\ 
D.O.F.		&1832				&1229		& 1591			&1464\\ 
$\log L_X$ &30.01 			&30.33		&30.31			&30.17 \\ 
$[0.2-10 \mathrm{keV}]$ &&&&\\ 
\hline 
\end{tabular}  
\end{center} 
\end{table}

\subsubsection{The FIP effect}\label{fip1} 
 
\begin{figure} 
\begin{center} 
\includegraphics[width=8.0cm]{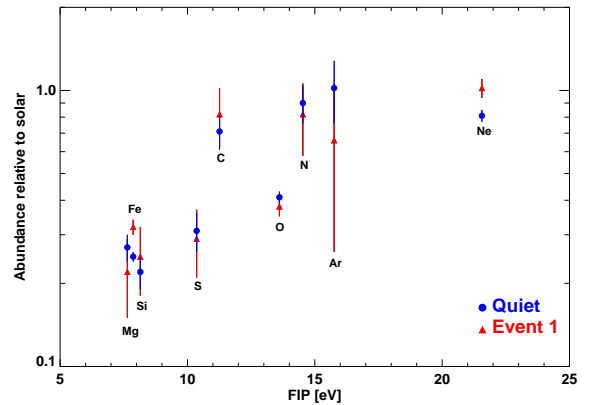} 
\caption{\label{fip}  AB~Dor~A's coronal abundance relative to solar photospheric values \citep{Grevesse} as a 
function of the first ionisation potential (FIP) during quasi-quiescence (blue) and Event 1 (red).} 
\end{center} 
\end{figure}

The measured abundance patterns of AB~Dor~A during flare and quiescence are  
shown in Fig. \ref{fip}, 
where we plot the abundances with respect to solar photospheric 
abundances against the FIP (first ionisation potential) 
of the corresponding element. 
Inactive stars like the Sun show the so-called `FIP effect', where low-FIP elements  
like Fe, Si, Mg etc. are  
enhanced in the corona when compared to high-FIP elements like C, N, O, Ne, etc..  
However, a reverse pattern called the `inverse FIP effect' (IFIP)  is observed  
in active stars (e.g. 
\citealt{Brinkman}, \citealt{Audard}). As can be seen in Fig.~\ref{fip}, 
the abundance pattern of AB~Dor~A for both quiescent and flaring state  
indicates the  
inverse FIP effect, which is consistent with the results of \cite{guedel}.  
 
\subsubsection{Coronal densities} 
\label{densitiesne} 
 
\begin{figure} 
\begin{center} 
\includegraphics[angle=0,width=7cm]{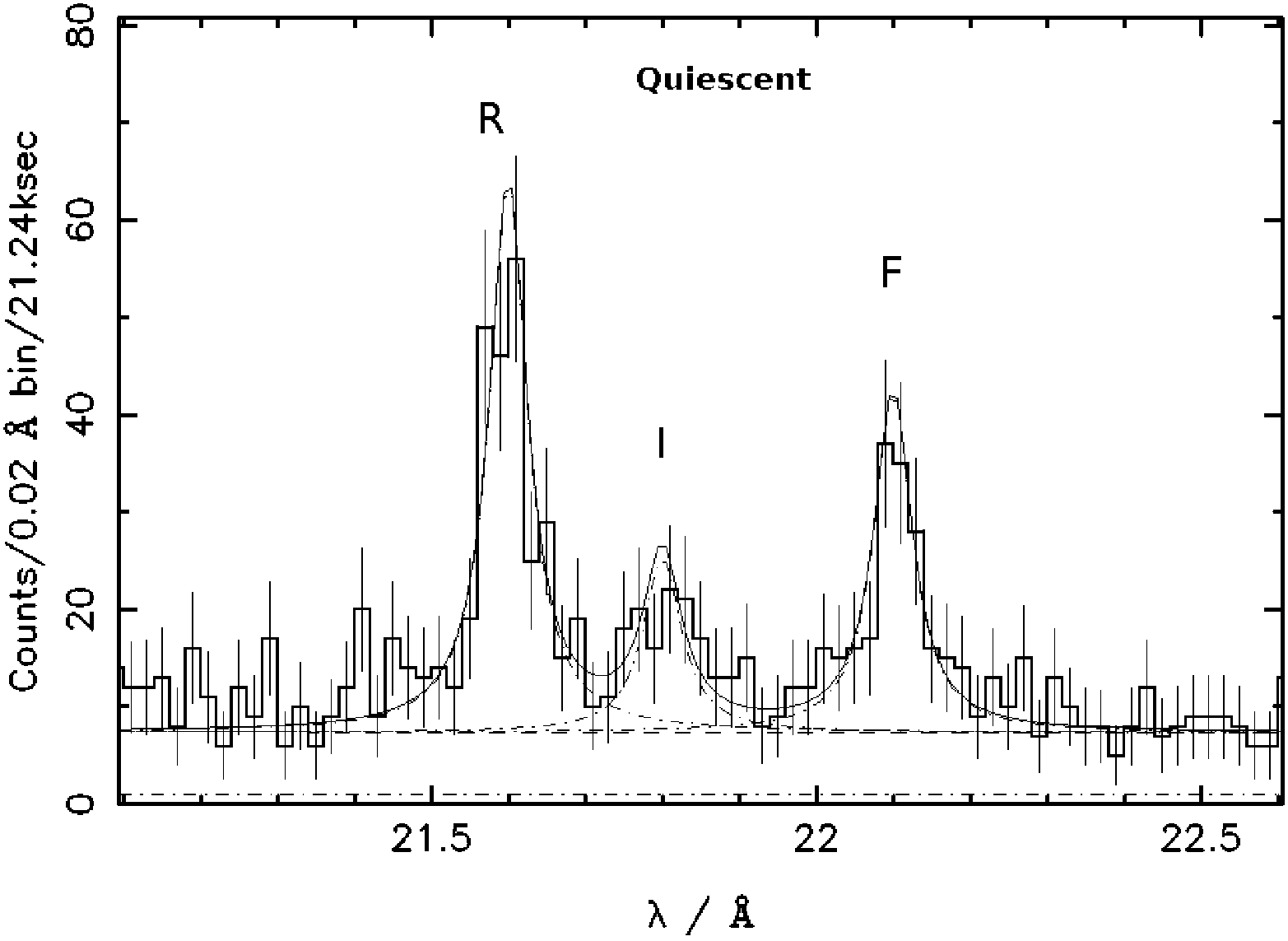} 
\includegraphics[angle=0,width=7cm]{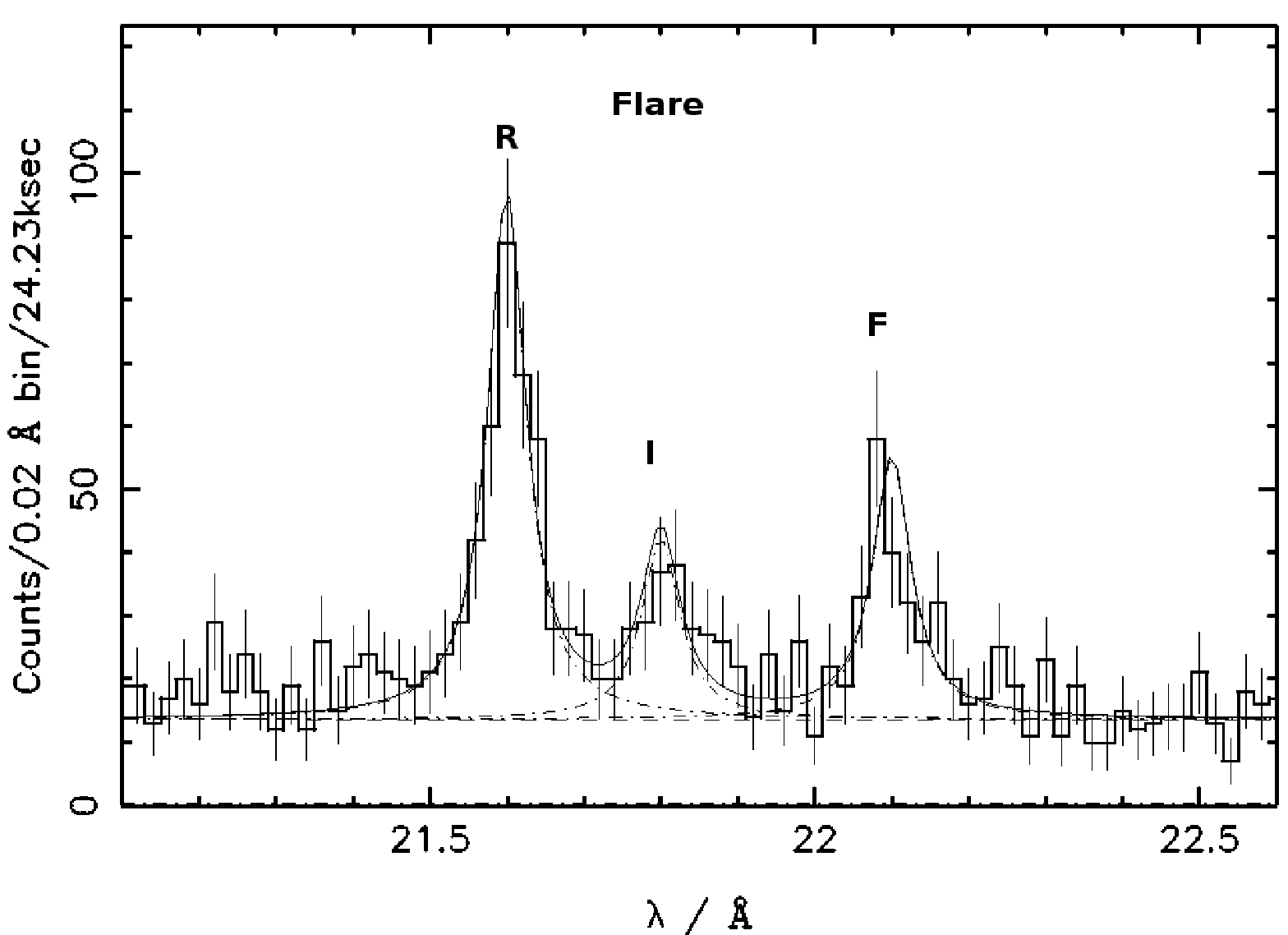} 
\caption{\label{density}The density-sensitive He-like triplets of \ion{O}{vii}  
(resonance, inter combination, and forbidden line for increasing wavelength) measured with the RGS1 with best fit Lorentzians  
during preflare quiescence (top) and flaring (bottom) state. Represented as dotted-dashed lines are the three Lorentzians representing r, i and f lines.  
The continuum close to the triplet is approximated by a straight horizontal line.} 
\end{center} 
\end{figure}

Using our RGS spectra we can investigate the electron densities of the coronal plasma from an analysis of the density-sensitive line ratios of  
forbidden to inter-combination lines of the 
helium-like triplets (\ion{N}{vi}, \ion{O}{vii}, \ion{Ne}{ix}, \ion{Mg}{xi}, and \ion{Si}{xiii}); 
the theory of density-sensitive lines has been described in detail by \citet{Gabriel_Jordan}. 
Only the He-like triplet of \ion{O}{vii} is strong enough 
to be used to obtain the characteristic electron densities in the 
source region. In Fig.~\ref{density} we show the quiescent and flaring RGS spectrum of the \ion{O}{vii} triplet, together with the best fits to the triplet 
lines r (resonance), i (inter-combination), and f (forbidden) provided 
by the CORA program \citep{cora}. 
The measured line counts and the deduced $f/i$ ratios are listed in Table~\ref{oviitable} 
for the pre- and post-flare quiescence (integration time of 20~ks and $\sim$10~ks, respectively), 
the individual and summed flaring events. 
 
\begin{table*} 
\caption{\label{oviitable} X-ray counts measured by best fit to lines and $f/i$ ratios deduced 
from the \ion{O}{vii} triplet.} 
\fontsize{8.5pt}{10pt}\selectfont  
\begin{tabular}[htbp]{llllllll} 
\hline 
\hline 
line & preflare & Post-flare  & Event 1 & Event 2 & Event 3  & Total & All events\\
     & quiescence & quiescence & & &  & quiescence & \\
     & $\sim$21 ks & $\sim$9.8 ks  & $\sim$5.5 ks & $\sim$9.6 ks & $\sim$9.1 ks & $\sim$30.8 ks &$\sim$24.2 ks\\ 
\hline \\[-3mm] 
R~(21.6 \AA)	& 259$\pm$20 & 136$\pm$15  & 101$\pm$13 & 145$\pm$15 & 213$\pm$18 & 395$\pm$25 & 459$\pm$27\\ 
I~(21.8 \AA)	& 82 $\pm$14 & 41$\pm$10   & 43$\pm$10  & 62$\pm$12  & 71$\pm$13  & 123$\pm$18 & 175$\pm$20 \\ 
F~(22.1 \AA)	& 160$\pm$17 & 82$\pm$12   & 49$\pm$10  & 92$\pm$13  & 98$\pm$13  & 242$\pm$21 & 238$\pm$21 \\ 
\hline 
$f/i$  & 1.95$\pm$0.40 & 1.98$\pm$0.55  &$1.48 \pm 0.35 $ &1.38 $\pm 0.31$ & 1.36 $\pm$ 0.20 & 1.96 $\pm 0.32$ & 1.36 $\pm$ 0.19  \\ 
n$_{e}$ $[10^{10}$~cm$^{-3}$]& $3.17_{-1.05}^{+1.63}$ & $3.08_{-1.66}^{+1.74}$  &$5.17_{-1.59}^{+2.61}$ &$5.77_{-1.62}^{+2.57}$& $5.90_{-1.15}^{+1.55}$ &$3.12_{-0.87}^{+1.22}$ & $5.91_{-1.11}^{+1.45}$ \\ 

\hline 
Accounting for quiescence  \\
$f/i$ &  & &0.28$\pm$0.52 & 0.70$\pm$0.73 & 0.76$\pm$0.52& &0.57$\pm$0.39\\
n$_{e}$ $[10^{10}$~cm$^{-3}$] & &  &$40.35\pm78.18$ &$14.27\pm14.92$ &$12.87\pm7.80$& &$18.35\pm11.92$\\ 

\hline
\end{tabular} 
\label{tab3}
\end{table*} 
 
\begin{figure*} 
\begin{center} 
\includegraphics[width=9.5cm,height=6.1cm]{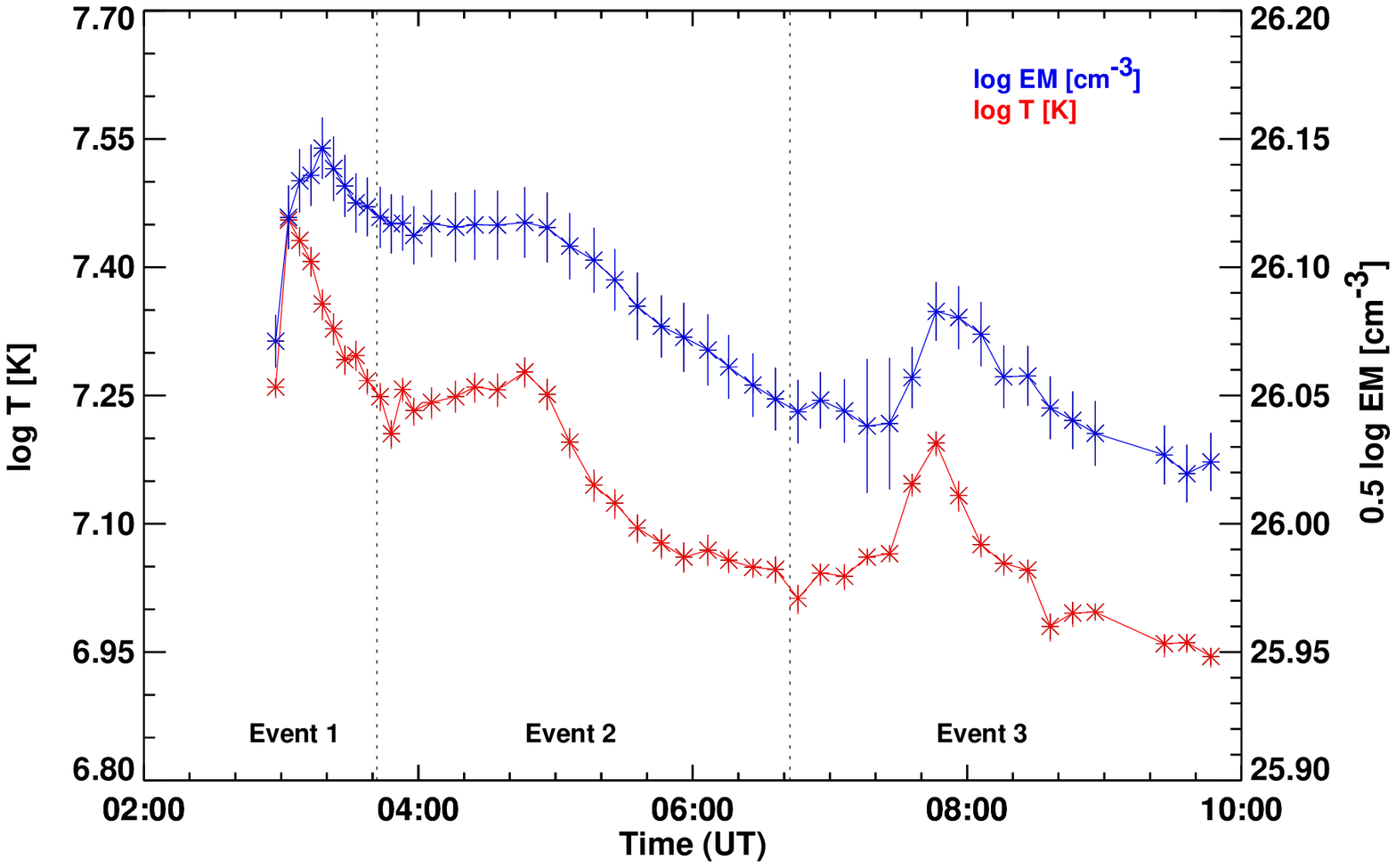} 
\includegraphics[width=8.5cm, height=6cm]{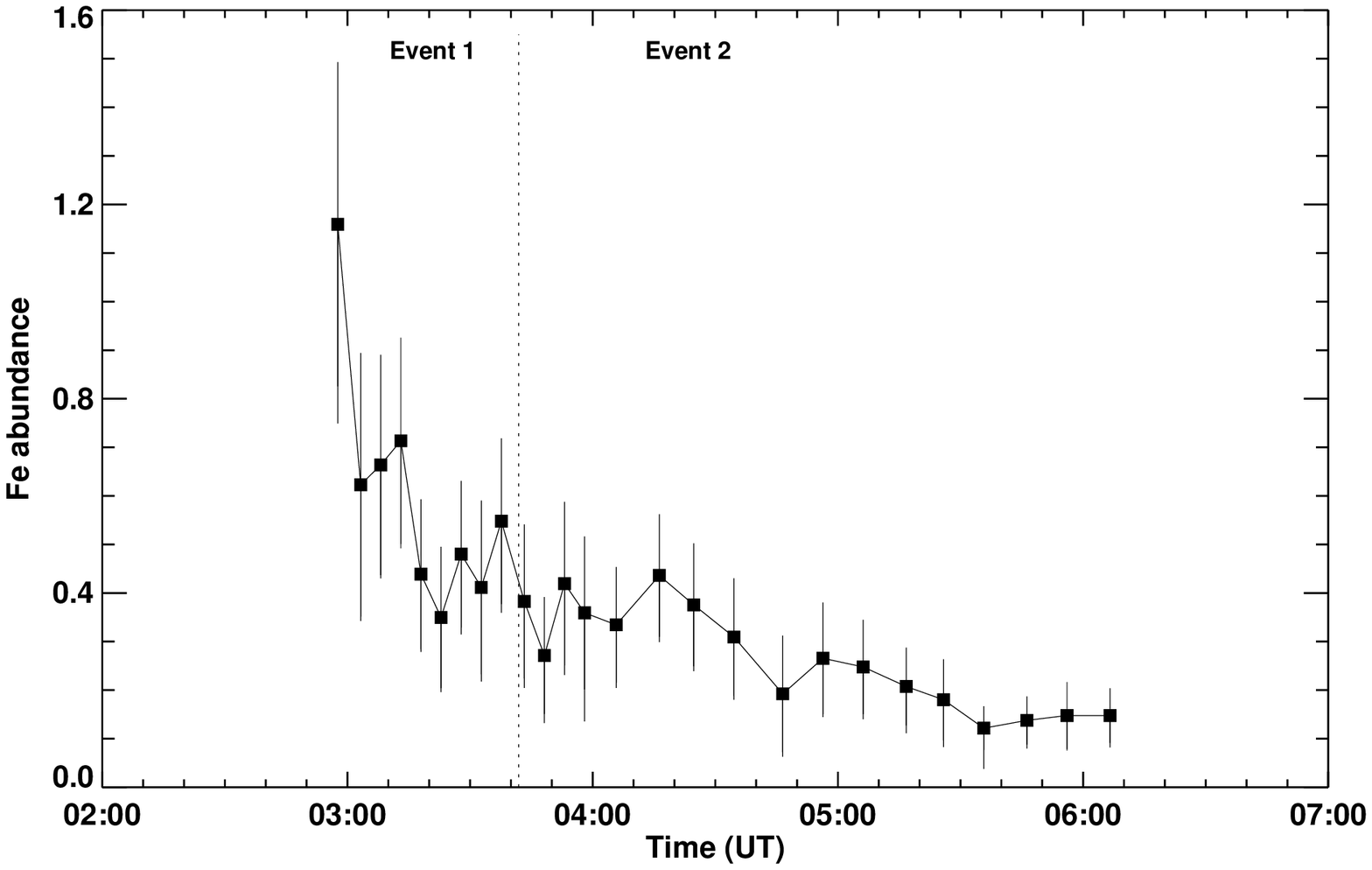} 
\caption{\label{em_evol_temp} Temporal evolution of flare temperature,  
emission measure in (left panel), and Fe abundance (right panel). Dotted lines indicate the boundaries of each 
event.} 
\end{center} 
\end{figure*} 
 
To convert the measured f/i ratios to electron densities $n_{e}$, 
we use the expression 
 \[ 
\frac{f}{i} = \frac {R_{o}}{1+\frac{n_{e}}{N_{c}}},
\] 
where $R_{o}$ denotes  the low density limit and $N_{c}$  the critical density and
adopted the values of $R_{o}$=3.95 and $N_{c}$=3.1 $\times$ 10$^{10}$~cm$^{-3}$ from \citet{Pradhan}.   
The line ratio errors are large because of the weakness of the inter-combination line. 
Since the errors overlap, there is -- formally -- no significant change in density.  
However, the above density estimates for the flares have not been corrected for 
any contribution from the quiescent emission.  
If the quiescent emission contribution is accounted for in the flare data, by 
subtracting the individual line fluxes scaled by the
respective exposure times, then the f/i-ratios  
decrease further (cf., 6$^{th}$ row in Tab.\ref{tab3}). Indeed, the actual flare plasma densities 
becomes very high though it should be noted that the associated 
measurement errors are even higher.
 
\subsubsection{Emission measure, temperature, and iron abundance evolution}\label{emtfe_evolve} 
 
In our previous investigation we only considered the integrated flare spectra. 
To investigate the variations in temperature, the emission measure, and 
abundance variations in more detail,  
we divided the data covering Events 1 to 3 into several time intervals and 
created X-ray spectra for each of these intervals. 
The first spectrum covers the flare rise, and the  
following time intervals cover the different phases of the decay by employing
300s and 600s bins so that they  
contain an approximately equal number of counts per spectrum.  
Each of these spectra was fitted using a superposition of four-temperature APEC models.
Our models always include quiescent emission; i.e., the first two temperature components  
and its parameters are fixed to plasma properties of the quiescent spectrum  
before the flare. With this approach, we account for the contribution 
of the quiescent coronal plasma to the overall X-ray emission. The third and fourth 
temperature components are allowed to vary freely, but we keep the elemental abundances fixed to quiescent values, 
except for the iron abundance, 
which was allowed to vary freely (but of course identical in both the temperature components).

To compare the plasma properties of the two temperature components accounting for the flare, 
we calculate the
``total EM'' as the sum of their emission measures
and a ``flare temperature'' as an emission measure-weighted 
sum of the two temperatures.
In Fig. \ref{em_evol_temp}, we plot the evolution of the flare temperature, 
the total emission measure and the Fe abundance. Owing to poor count statistics we end Fig.~\ref{em_evol_temp} before the onset of Event~3. 
The temperature and the emission measure evolution exhibits a 
decay-like light curve, and the iron abundance also increases from the  
quiescence level directly before the flare to a maximum during the flare peak  
(Fig. \ref{em_evol_temp} right panel) and then decreases to preflare values. 
The increase in Fe abundance around 4:20~UT may not be significant, nevertheless,  
it coincides with the second peak of the hardness ratio light curve which we define as 
Event~2 (see Fig.~\ref{light_curve}). For Event 3 we also checked for iron 
abundance changes, but because of lower S/N in the data we find no significant 
change in iron abundance compared to the quiescent state. 
 
Clearly, the results in Table~\ref{obs} suggest that both the Fe and Ne 
abundances vary significantly. We therefore experimented with various fit strategies 
and found that the neon abundance does not show a clear pattern like the 
iron abundance in  Fig. \ref{em_evol_temp} (right panel), when it is allowed
to vary freely. Furthermore, the fits are only marginally improved, and most
of the fit residuals remain unchanged.  We also linked Fe and Ne elemental 
abundances and allowed them to vary freely, but no improvement in the fit 
quality of the fit was noticed when compared to the fits where only Fe abundance is set free alone. 
Finally, we repeated the fits treating oxygen, silicon, and their combination as free parameters,
but this strategy leads to poorly defined abundances for the
individual spectra.  Therefore, Fe seems to be the only element that shows a clearly  
measurable abundance variation during the flare. We cannot exclude the
variations of other elemental abundances as well, however, given the spectral
resolution and the S/N ratio of our time-resolved X-ray spectra. Such variations -- if present -- remain hidden in the noise.

\subsection{Loop properties}\label{loopEMT}

Observations of solar and stellar flares have  
shown a correlation between the emission measure and the peak flare
temperature. The so-called emission measure - temperature (EM-T) diagram 
has become a useful diagnostic for estimating physical quantities that are not directly observable. 
By computing EM-T diagrams from specific physical models
and comparing and fitting observed and theoretical EM-T curves, one can infer
the physical properties of the flare within the chosen model context.
With the data from our multi-wavelength campaign, we can assess whether the chosen
model is appropriate. Here the model assumes that a stellar flare occurs  in a localised coronal region in a simplified 
geometry (i.e., single loop structure), which remains unchanged during the 
flare evolution. While the heating is a free parameter in these models, plasma
cooling occurs by radiation and thermal conduction back to the chromosphere 
with their characteristic cooling time scales. These cooling time scales
depend on the size of the confining loop structure and on the density of
the radiating plasma \citep{Reale}.  Specifically,
\cite{serio} derive a thermodynamic decay time scale of a flare confined to a 
single flaring coronal loop, by assuming a semicircular loop with a constant cross section, 
uniformly heated by an initial heating event and without further heating during the decay phase. 
In this case the loop thermodynamic decay time scale is given by  
\[ 
\tau_{s} = \alpha \frac{L}{\sqrt{T_{0}}} ~ = 120 \frac{L_{9}}{\sqrt{T_{0,7}}}, 
\] 
\noindent
where $\alpha=3.7 \times 
 10^{-4}$ cm$^{-1}$s\,K$^{1/2}$, $T_{0}$ ~(T$_{0,7}$) is the loop maximum temperature in units of $10^7$ K, 
 and L~($L_9$) is the loop half length in units of $10^9$ cm.

To account for further heating during the flare decay,
\citet{2007} derived an empirical correction factor by a hydrodynamic 
simulation of the loop decay. In this case the loop length can be estimated by the 
formula
 
\begin{equation}
 L  = \frac{\tau_{LC} \sqrt{T_{0}}}{\alpha F(\zeta)} ~~ or ~~ L_{9}= \frac{\tau_{LC} 
 \sqrt{T_{0,7}}}{120 F(\zeta)} ~~~~~  \zeta_{min}<\zeta\leq\zeta_{max} ~ , 
\label{reale}
\end{equation} 
\noindent
where $\tau_{LC}$ denotes the decay time derived from the light curve.  Here
the unitless correction factor F($\zeta$) describes the ratio of the observed and the thermodynamic decay time; 
F($\zeta$) is approximated with an analytical function as  
 $F(\zeta)= \frac{\tau_{LC}}{\tau_{th}}=\frac{c_{a}}{\zeta - \zeta_{a}} + q_{a}$  
where $\zeta$ is the slope of the   
EM-T diagram. The observed peak temperature must be corrected to 
 $T_{0} (T_{0,7}) = \xi T_{obs}^{\eta}$ (in units of $10^{7}$ K), where $T_{obs}$  
is the best-fit peak temperature derived from spectral fitting to the data. 
The coefficients  $\xi$, $\eta$, $c_{a}$, $\zeta_{a}$, and $q_{a}$ depend on 
the specific energy response of the  
instrument used: for XMM/EPIC the appropriate values are $\xi = 0.13$, 
 $\eta = 1.16$, $c_{a} = 0.51$, $\zeta_{a} = 0.35$, and $q_{a} = 1.35$ (cf. \citep{2007}).

We apply Eq.\ref{reale} to each of the three events. 
We disentangle the events by modelling them assuming an exponential growth/exponential 
decay, given by $CR= CR_{flare} \times exp\pm\frac{t-t_{flare}}{\tau_{decay}}$ where 
$CR_{flare}$ is the count rate at flare peak,  
$t$ and $t_{flare}$ are the observation time and the time of flare peak, respectively.

\begin{figure}[h] 
\begin{center} 
\includegraphics[width=8cm]{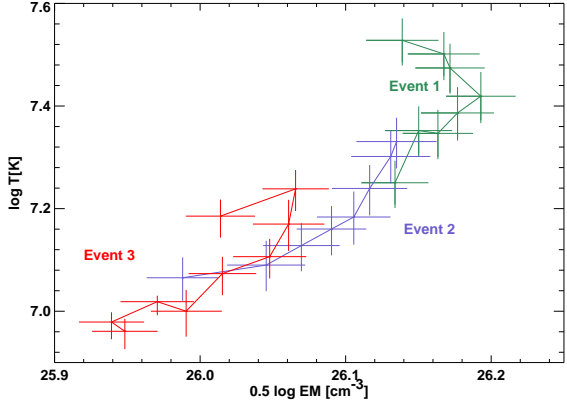} 
\caption{\label{em_t} Flare evolution in density and temperature for Event 1 (green), Event 2 (blue), and Event 3 (red).} 
\end{center} 
\end{figure}

In Fig. \ref{em_t} we show the evolution of the three events in the EM-T plane, which we use to measure the slope 
$\zeta$ of the EM-T evolution.  Using these values,
the temperatures derived from the spectral fits corrected using the correction factors given in \citet{2007}, 
as well as the decay times derived from the light curve,
we can determine the loop half lengths $L$ by using Eq.\ref{reale}; the input parameters
used and the derived loop lengths are provided in Table \ref{loop_table}. The length
scales listed in Table \ref{loop_table} for Events~1 and 3 suggest that these 
flaring structures have lengths
of 1-2 $\times$ 10$^{10}$cm$^2$, while the derived loop length for Event~2 is
significantly larger. 
However, this value is presumably overestimated for Event~2 since 
additional heating seems to occur during its decay. 
In that case, Event~2 would not comply with the model of a simple, single flaring 
loop used in the \citet{Reale} model.

\begin{table} 
\caption{\label{loop_table} Decay time $\tau_{LC}$ derived from the light curve, slopes $(\zeta)$ from the EM-T diagram,  
flare peak temperature T, and  
deduced loop lengths L for each of the events.} 
\begin{tabular}[htbp]{lrrr} 
\hline 
\hline 
Parameters & Event 1	 & Event 2     & Event 3\\ 
\hline \\[-3mm] 
$\tau_{LC}$ [ks]		& $0.94\pm0.02$ & $7.72\pm0.03$ & $1.90\pm0.01$ \\ 
$\zeta$	[slope in $Kcm^3$]	& 1.67$\pm$0.05	 & 1.90$\pm$0.12 & 1.80$\pm$0.12\\ 
T [MK] & $70.21\pm1.29$ & $65.79\pm1.31$ & $55.43\pm1.85$\\ 
L [$10^{10}cm$] 		& 1.22$\pm$0.02 & 8.86$\pm$0.05 & 2.21$\pm$0.04\\ 
$\frac{L}{R_{star}}$ & 0.18$\pm$0.01 & 1.33$\pm$0.08 & 0.33$\pm$0.02\\
\hline 
\end{tabular} 
\end{table}

\section{Chromospheric properties of AB~Dor~A}\label{chrom_prop} 
 
\begin{figure*} 
\begin{center} 
\includegraphics[width=16cm]{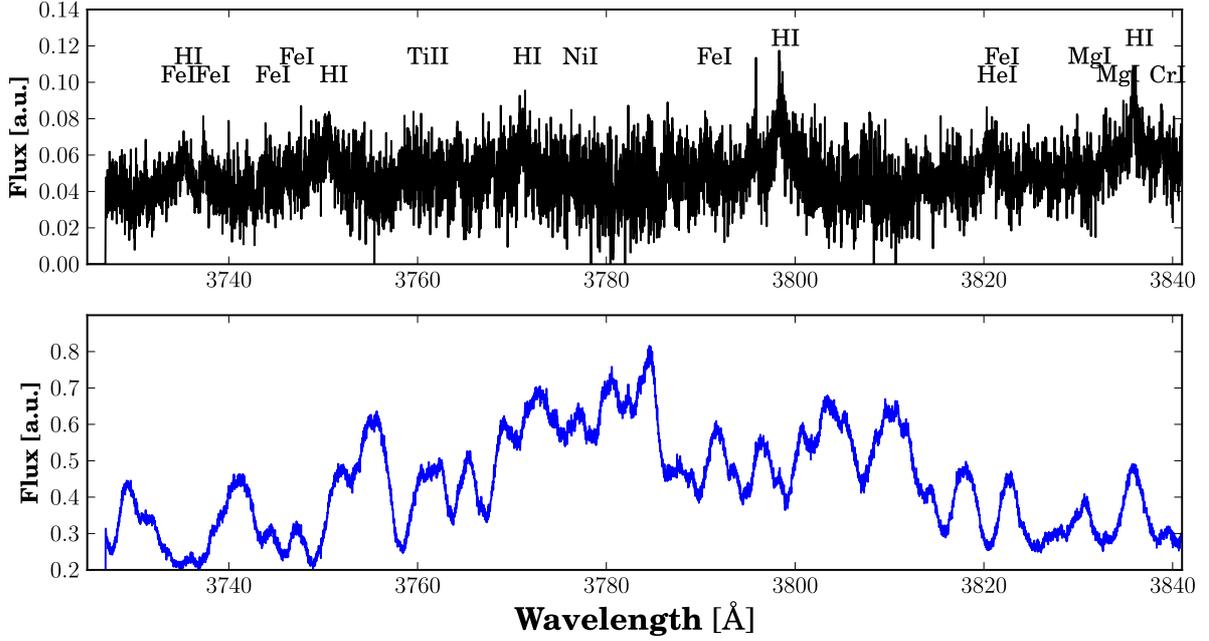} 
\caption{\label{blue_spectrum} Flare onset spectrum of AB~Dor~A in an arbitrary chosen wavelength range covering 
the blue end of our observations including the Balmer lines H9 -- H13. 
 Top: The flare spectrum with a quiescent spectrum subtracted. The 
identified emission lines are labelled. Bottom: The original flare spectrum.} 
\end{center} 
\end{figure*}

Optical lines, sensitive to chromospheric activity, are found throughout our 
UVES spectra, most of them in the blue wavelength range up to about 4000 \AA. 
 For AB~Dor~A the strongest chromospheric lines; i.e., the 
Balmer lines, \ion{Ca}{ii} H \& K, and \ion{Na}{i} D lines also show variations outside 
of flares, but only during flares do shallow metallic emission lines appear. 
To illustrate that effect for Event 1, in Fig. \ref{blue_spectrum}, we show an arbitrarily chosen wavelength range for 
both the original recorded
spectrum (lower panel) and the net flare spectrum with the 
quiescent spectrum subtracted. We note that these emission lines can only be 
recognised in the subtracted spectra 
in contrast to active mid M-dwarfs, such as Proxima Centauri or CN Leo, where the  
quiescent continuum in the blue wavelength range is so weak, that the emission lines easily outshine 
the continuum, especially during flares.  
Also, the relatively strong photospheric continuum explains  
why in comparison to these very late-type  stars,  
far fewer emission lines are found in the spectrum of AB~Dor~A \citep{CNLeoflare,Proxcen}. 
Moreover, the high rotational velocity of AB~Dor~A broadens the lines leading to a lower amplitude, 
on the other hand, the rotationally broadened  lines offer the 
possibility to identify emission of single active groups, if strong lines show subcomponents.

In the following we first discuss the evolution of strong lines including  
the absorption features caused by prominences. Moreover, we present an emission line 
catalogue for events 1 and  3. Then we concentrate on the analysis  of 
some selected lines. 
 
\subsection{Chromospheric line evolution overview} 
 
\begin{figure*} 
\begin{center} 
\includegraphics[width=17cm]{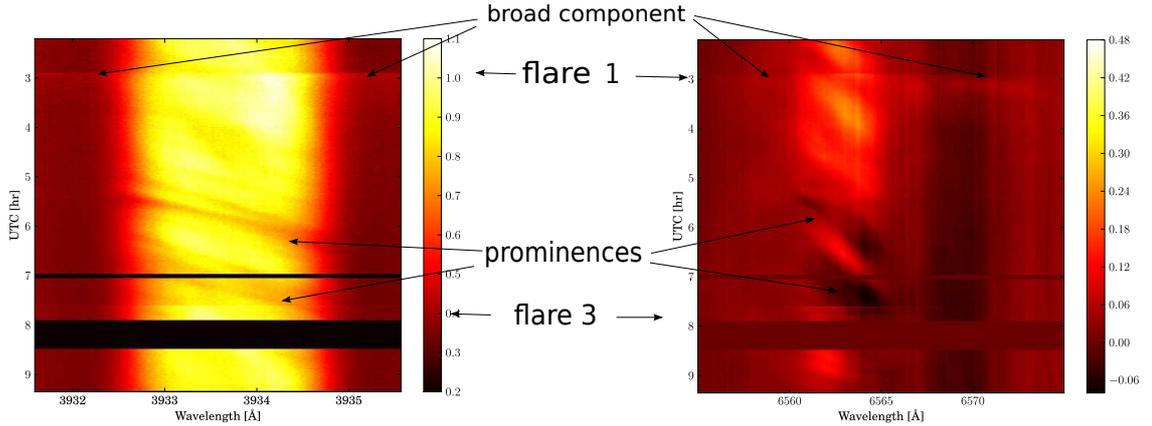} 
\caption{\label{intensitymap} Intensity map of the H$\alpha$ line (right -- with a PHOENIX  
photospheric spectrum subtracted) and for the \ion{Ca}{ii} line (left -- original 
spectrum, no spectrum subtracted). The two horizontal bars indicate observational gaps.} 
\end{center} 
\end{figure*}

As an example of the development of an emission line throughout our observations we show 
an intensity map for the H$\alpha$ and the \ion{Ca}{ii}~K lines in Fig. \ref{intensitymap}. 
For the H$\alpha$ line we subtracted a PHOENIX spectrum ($T_{\rm eff}$ =4900 K and log g=4.5, for more details see below), 
while we show the original spectrum for the \ion{Ca}{ii} K line. 
The main flare onset can be noted as a bright horizontal line directly before 3:00 UT. 
For both lines one can see the line broadening directly after flare onset: for the H$\alpha$ 
line, there is a significant brightening on the red side up to 6575 \AA, while 
for the \ion{Ca}{ii} K line the core is broadened. The  
\ion{Ca}{ii} K line shows its strongest brightening directly at flare onset. In contrast, the 
H$\alpha$ line shows the strongest brightening about half an hour later, and it seems to be 
drifting from blue to red, making it questionable that the brightening is really related to the main flare. 
Also the third event shows up in the intensity map, starting at about 7:30 UT, directly before 
the second observation gap and extending to the end of the observations.

Judging from the H$\alpha$ line and compared to previous observations, 
AB~Dor~A is in a state of medium activity during our whole observation. 
The absorption transients in the original spectra reach a depth of about 0.8-0.85, which seems to 
be quite typical compared to the findings of \citet{Collier_Cameron} and \citet{vilhu2}. 
The latter also detected significant H$\alpha$ emission in the unsubtracted spectra, which they ascribe to flaring activity or a high 
density prominence. 
The flaring activity in our observations does not exhibit such a pronounced H$\alpha$ emission  
line, consistent with our X-ray observations, which 
would also suggest a medium activity state during our observations. 
 
\subsubsection{Crossing prominences}
\label{prominence}  

In addition to the brightening events a few dim structures can be noted in 
the intensity map between 5:00 and 7:30 UT. 
Such absorption transients are frequently observed
in the spectra of AB~Dor~A and some other highly active stars. 
These absorption features drift {\it very quickly}  
across the line profiles of the Balmer and \ion{Ca}{ii} H\&K lines; 
occasionally, these lines even show such transients in emission. 
Following early analyses of this phenomenon \citep{Collier_Cameron} 
these transients are usually ascribed to clouds of circumstellar material 
held -- at least to a large extent -- in co-rotation with the star 
by strong coronal magnetic fields. 
Interpreting the drift velocities of these transients as caused by rigid rotation
with the stellar surface leads to remarkably large elevations of the putative 
co-rotating clouds above the stellar surface, reaching heights of up to about 
five stellar radii or even more \citep{Dunstone, Wolter}. 
 
Assuming a stellar inclination of \mbox{$i=90\degr$}, the velocity of a 
structure at a distance $r$ from the  stellar rotation axis projected 
onto the line of sight is given by \mbox{$v_r = \omega r \sin \omega t$}, 
where $\omega$ denotes the angular velocity of the star.
The observed wavelength of the transient is determined by this velocity,
which exceeds the rotational velocities on the stellar surface. 
In our case, the curvature of the line profile transients,  
\mbox{$\frac{d^2}{dt^2} v_r$}  
cannot be determined. As a result, we approximate the above expression 
to first order. Differentiation yields  
\mbox{$\dot v_r = \frac{d}{dt} v_r = \omega^2 r$}; and finally, 
including the projection due to the stellar inclination, we obtain 
\begin{equation} \label{rad}
 \qquad r = \frac{\dot{v_r}}{\omega^2 \sin i},
\end{equation} 
which allows us to compute the height $r$ for a line profile transient  
as a function of its observed ``drift velocity'' 
$\dot v_r$ through the spectrum. 
We note that $r-R_\star$ is the height above the stellar surface, 
{\it only} if the absorbing cloud is located in the equatorial plane. 
In general, the determined value of $r-R_\star$ is a {\it lower} limit to its height.
To convert the values of r into stellar radii, we used the 
\mbox{$R_\star=0.96\pm0.06 \, R_{\sun}$}, as determined by 
\cite{guirado}   
based on VLTI/AMBER interferometry in the NIR, 
and an adopted distance of \mbox{$14.9\pm0.1$~pc}.

As can be seen in Fig. \ref{intensitymap}, \ion{Ca}{ii} K shows sharper  
transients than H$_\alpha$. On closer inspection both lines show multiple 
transients. The properties of the four main transients that could be identified in both lines, 
\ion{Ca}{ii} K and H$_\alpha$, are listed in Table~\ref{prominences}. We give the approximate 
time for crossing the line centre, so that the prominence can be identified in Fig. \ref{intensitymap}. 
 Moreover, we list the drift velocity and the height above the 
stellar surface using an inclination of 60$^\circ$ \citep{kuerster_1994,donati_2003}.

To quantify the uncertainty of the resulting heights, we estimated the 
typical error of our determined drift velocities to 15 km/s/h, i.e. about 10\%. 
The stellar inclination is best determined by Doppler imaging, its precise uncertainty 
is difficult to assess. Looking at Fig.~3 of \citealt{kuerster_1994}, it can be determined 
within a margin of 10 to 15 degrees. This amounts to an uncertainty of about 15\% in $\sin i$.
Plugging these numbers into Eq.~\ref{rad} and neglecting the small 
uncertainty of the stellar rotation period, 
the relative uncertainties of the drift velocity and $\sin i$ approximately add up 
to a relative uncertainty of 
25\% for the resulting prominence radii. A formal error propagation leads to a 
similar result. 
Additionally, it should be kept in mind that the heights given in 
Table~\ref{prominences} are lower limits to the 
true prominence heights.
The actual heights depend on the prominence latitudes which, though
apparently close to the equator, 
are not known in detail. 
\begin{table} 
\caption{\label{prominences}Properties of the main transients.} 
\begin{tabular}[htbp]{cccc} 
\hline 
\hline 
Time  & drift velocity  & distance r & height $[R_\star]$ \\ 
$[$ UT $]$& $[$ km\,s $^{-1}$\,h$^{-1}]$&  $[$ m $]$ & \\
\hline
5:20 & 140  & 2.3$\times 10^{9}$& 2.5 \\ 
5:40 & 175  &  2.9$\times 10^{9}$& 3.3 \\ 
6:30 & 150  & 2.5$\times 10^{9}$& 2.7 \\ 
6:50 & 160  & 2.6$\times 10^{9}$& 3.0 \\ 
\hline 
\end{tabular} 
\end{table}

In addition to the radial extent of the tentative prominences 
it would be interesting 
to estimate their lateral size relative 
to the stellar disk, following the procedure of  
\cite{Cameron-1990}. 
However, identifying an actual feature 
belonging to a given transient in the individual line profiles   
turned out to be cumbersome -- owing to the multitude of significant features present in 
our spectra. 
Thus, we consider such an analysis beyond the 
scope of this paper.

We also searched for X-ray absorption features at the prominence crossing times, 
but could not find any evidence of an X-ray imprint of these features. Using the X-ray data we estimate 
an upper limit of N$_H$=9$\times 10^{18}$ cm$^{-3}$ to the hydrogen column density 
of the prominence material. 
While this number implicitly assumes cosmic abundance for the absorbing material, however, 
the quoted values are independent of the actual ionisation state 
of hydrogen.

\subsection{Line fitting procedure} 
 
In order to quantitatively determine
the properties of individual chromospheric emission lines, we performed fits 
using up to three Gaussian components treating the width $\sigma$, the amplitude, 
and central wavelength as free parameters for each component. Since we are dealing 
with 460 spectra, the lines were fitted semi-automatically; i.\,e., we checked the 
fit quality by eye for each spectrum and line, and if the fit did not reproduce the 
observed line profile well, we changed starting parameters and/or the fit parameter restrictions. 
Often the line shapes are quite complicated, such as when the \ion{Na}{i} doublet being an 
absorption line with some filling in, but without any emission core, which leads to
poor fits in the context of our simple fitting approach.
We checked the possibilities 
of subtracting either an observed quiescent spectrum or a PHOENIX photospheric model spectrum \citep{phoenix}. 
For the latter we determined the best-fitting PHOENIX spectrum for different quiescent 
spectra of AB~Dor~A using a model grid with $T_{\rm eff}$  ranging from 4700 to 5200 K 
in steps of 100 K, with log g ranging from 3.5 to 5.0 in steps of 0.5, and with 
the rotational velocity ranging from 60 to 120 km\,s$^{-1}$ in steps of 10 km\,s$^{-1}$ using 
solar metallicity. Our best fit values are a $log\  g$ of 4.5, an effective temperature
of 4900 $\pm 100$ K, and a projected rotational velocity of  
100 $\pm$ 10 km\,s$^{-1}$,  
in agreement with the value of 91 km\,s$^{-1}$ adopted by \citet{Jeffers}. 
The PHOENIX model spectrum describes the photospheric lines quite well in general, although 
the amplitudes of individual lines differ in many cases.  
 
As individual lines we fitted the \ion{Ca}{ii} K line (the H line is too complicated 
due to blending with Balmer line emission), the \ion{Na}{i} doublet at 5889 and 5895~\AA, 
the \ion{He}{i} D$_{3}$ line, H$\alpha$, H$\beta$, H$\delta$, and, as an example of a shallow 
metal line, the \ion{Si}{i} line at 3905~\AA. The \ion{Si}{i} and the \ion{He}{i} D$_{3}$ line 
were both fitted using a running mean; i.\,e., for each fit, two consecutive spectra 
were averaged, nevertheless moving through every spectrum.

\subsubsection{Observed vs. simulated spectrum as quiescent template}\label{obsvssim} 
 
As shown in Fig. \ref{blue_spectrum}, a quiescent template spectrum must be subtracted from 
the flaring spectra in order to identify and properly analyse the flare emission lines.  
Using a model spectrum  or an observed spectrum as quiescent standard has  
both advantages and disadvantages.  
 
When subtracting a PHOENIX model spectrum, the  
deviations of the model from the observed photosphere may be as large as the amplitudes 
of the low amplitude metallic chromospheric emission emerging during the flare. 
Therefore, a PHOENIX model can only be used for modelling
strong emission lines. For these lines the general chromospheric emission 
leading to the filling-in of the lines (or to emission cores) is often so large 
that after subtraction of a PHOENIX spectrum, one ends up with a strong emission line, 
with the flare and emission of single active regions only playing a minor role, which cannot be properly modelled with an automatic fit.  
This is only possible for the H$\alpha$ line, where the emission of the flare and 
individual components is strong enough to show up against the emission core.  
 
\begin{figure} 
\begin{center} 
\includegraphics[width=9cm]{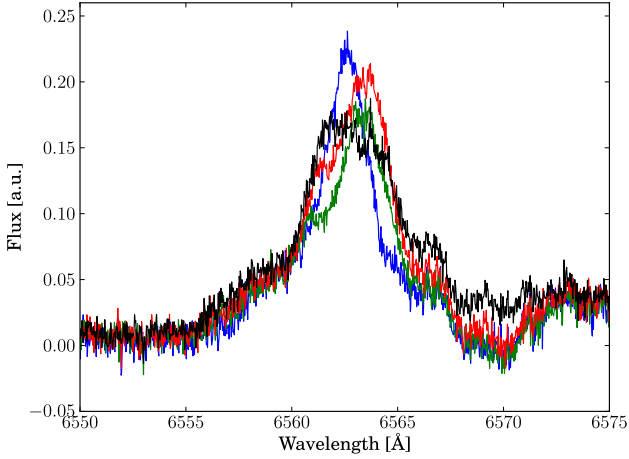} 
\caption{\label{halpha_phoenix}Examples from the evolution of the H$\alpha$ line after subtracting a PHOENIX  spectrum 
in the first hour of observations, covering the main flare. 
Blue: spectrum at 2:13 UT (first spectrum taken), green: spectrum at 2:48 UT; red:  
spectrum at 2:58 UT (flare onset); black:  
spectrum at 3:05 UT. The additional flux in the black spectrum (best seen at about 6557 and 6570 \AA) 
is caused by the broad line component.} 
\end{center} 
\end{figure}

Subtracting an observed non-flaring spectrum as a proxy for the quiescent spectrum 
has the disadvantage that the spectra change significantly on short time scales (about 10 minutes), 
with single features of the line moving either in central wavelength or changing in amplitude. 
For example, in the H$\alpha$ line, subtracting an average of the first three spectra 
from each single spectrum, after only about ten minutes of observations an additional emission 
feature is turning up, and again about ten minutes later, an absorption component 
manifests itself, but these features are only defined relative to the subtracted averaged 
spectrum. For example, the absorption feature is caused by the dimming and shifting of a strong   
emission component  in the averaged quiescent spectrum, and therefore not a `real' absorption component. 
This effect is illustrated for H$\alpha$ line in Fig. \ref{halpha_phoenix}, where the PHOENIX subtracted lines 
are shown and one notes the shifting and dimming. An example of an absorption line caused by changes 
in the quiescent emission can 
be found in Fig. \ref{halpha_diffspec} in the last two sub-frames. The distributed and 
fast changing chromospheric emission is also discussed in Sect. \ref{quasi_quiescence}.

Nevertheless, for several lines 
the fitting process is best done using an observed quiescent spectrum. This method gives 
reliable results for the lines that are not present before the flare or that show up only shortly before the flare  
(all lines and broad line components 
except the Balmer lines and the \ion{Ca}{ii} K narrow line components).

In the following sections we present  our findings from the line fitting. 
 
\subsection{Catalogue of emission lines}\label{catalogue} 
 
For Events 1 and 3, a number of chromospheric metal lines besides \ion{Ca}{ii} H \& K 
show up in the spectrum. Event 2 does not influence the shallow metal lines. 
Therefore, we compiled an emission line list for Events 1 and 3. 
For the line identification we generally used the Moore catalogue \citep{Moore}. 
The line fit parameters  -- central wavelength, half width, and flux in arbitrary units -- and   
(tentative) identifications including  some comments 
can be found in Table \ref{linetable}. We only show a few rows of this table  
in the paper as an example, while the whole table 
is provided  in electronic form at the CDS.\\
 
\subsubsection{Emission line properties}
 
For assessing the metal line properties for Event 1,  
we averaged the first and second flare spectrum  and subtracted an averaged quiescent spectra 
constructed out of the first three spectra of the night.
We identified a total of 90 emission lines in the blue arm and 11 lines in the red arm 
spectra with a mean wavelength shift of 39.6 $\pm$ 9.6 km\,s$^{-1}$ (applying a radial velocity of 32.5 km\,s$^{-1}$). The rather large scatter is caused by  
blended lines, for which the probable components are given as a comment in Table \ref{linetable}.  
Because of the high noise level, the line list cannot be considered complete with respect to weak emission features.
Most of the blue emission lines can be detected between 20 and 50 minutes after flare onset. Although 
Balmer line and \ion{Ca}{ii} H and K emission persists for about one hour after flare onset, this emission no longer appear to be flare-related. 
During Event 3 only 17 emission lines appear, and are a subset of the emission lines 
found during Event 1.  
Sufficiently strong lines significantly change in subsequent spectra, even in the relatively
weak Event 3. As for Event 1, 
we took the average of the three spectra directly preceding 
Event 3 as a proxy for a quiescent spectrum. 
The lines have an 
average line shift of $-23.9 \pm 13.2$ km\,s$^{-1}$, which differs by more than 
60 km\,s$^{-1}$ from the velocity measured for Event 1. 
 
\begin{table*} 
\caption{\label{linetable} Section of the line catalogue corresponding to the wavelength 
interval shown in Fig. \ref{blue_spectrum}. The whole 
table is accessible electronically. See Sect. \ref{catalogue} for details.}   
\begin{minipage}{\linewidth} 
\begin{tabular}[htbp]{cccclrc} 
\hline\hline 
 central wavelength & half width & flux & catalogued wavelength & ion & multiplet &  comment\\ 
$[\AA]$ & $[\AA]$ & a.u. & $[\AA]$    & & & \\ 
\hline \\[-3mm] 
3733.78 &   0.12  &   0.003 & 3733.319 &   FeI  &    5 &                      \\ 
3735.18 &   0.47  &   0.017 & 3734.370 &   HI   &    3 & blend with 3734.867    FeI     21\\ 
3737.42 &   0.13  &   0.007 & 3737.133 &   FeI  &    5 &                           \\ 
3743.80 &   0.15  &   0.005 & 3743.364 &   FeI  &   21 &                             \\ 
3746.32 &   0.27  &   0.005 & 3745.901 &   FeI  &    5 & blend with 3745.561    FeI      5\\ 
3750.59 &   1.06  &   0.042 & 3750.154 &   HI   &    2 & blend with 3749.487    FeI     21  \\ 
3759.88 &   0.11  &   0.005 & 3759.291 &   TiII &   13 & \\ 
3771.09 &   0.45  &   0.026 & 3770.632 &   HI   &    2 & \\ 
3776.35 &   0.13  &   0.006 & 3775.572 &   NiI  &   33 & possible blend, not in ProxCen list, but in CNLeo list\\ 
3790.58 &   0.10  &   0.005 & 3790.095 &   FeI  &   22 & not in ProxCen list, but in CNLeo list\\ 
3798.44 &   0.43  &   0.030 & 3797.900 &    HI  &     2&   \\ 
3820.25 &   0.09  &   0.005 & 3819.606 &   HeI  &   22 & blend with FeI 20 at 3820.428 \\  
3821.03 &   0.32  &   0.020 & 3820.428 &   FeI  &   20 & blend with HeI 22 at 3819.606 \\ 
3829.88 &   0.05  &   0.002 & 3829.355 &   MgI  &    3 & \\ 
3832.89 &   0.09  &   0.003 & 3832.303 &   MgI  &    3 & \\ 
3835.95 &   0.32  &   0.029 & 3835.386 &   HI   &    2 & \\ 
3836.52 &   0.25  &   0.008 & 3836.070 &   CrI  &   70 & not in ProxCen and in CNLeo list \\  
3838.86 &   0.07  &   0.002 & 3838.294 &   MgI  &    3 & \\ 
\hline 
\end{tabular} 
\end{minipage} 
\end{table*}

\subsubsection{Comparison to other line catalogues} 
 
We compared the emission lines found in our AB~Dor~A spectra to those found in Proxima Centauri by 
\citet{Proxcen}, which have an overlapping wavelength region between 3720 and 4485 \AA\, and  
between 6400 and 9400 \AA.  
We found that most of the lines in the AB~Dor~A flare coincide with the strongest 
lines identified in the Proxima Centauri flare. Nevertheless, there are 21 lines 
in the AB~Dor~A flare not found in the Proxima Centauri flare (see 
remarks in the line table). We therefore compare the line list also to 
the one of the CN~Leo mega-flare 
\citep{CNLeoflare}, where the wavelength overlap is, however,  much smaller. 
In the overlapping line region, all lines not found in the Proxima Centauri list 
could be identified except for one line.

\subsection{Timing behaviour of the strongest lines} 
 
The flux in the three strong lines H$\alpha$, H$\beta$, and less pronounced for  
\ion{Ca}{ii} K shows a slow increase and (later on) decrease well before Event 1 
(see Fig. \ref{mainflux}; the H$\alpha$ and \ion{Ca}{ii} K fluxes peak at about 2:30 UT, while the H$\beta$ flux 
peaks at about 2:40 UT).

\begin{figure*}[!ht] 
\begin{center} 
\includegraphics[width=5.8cm]{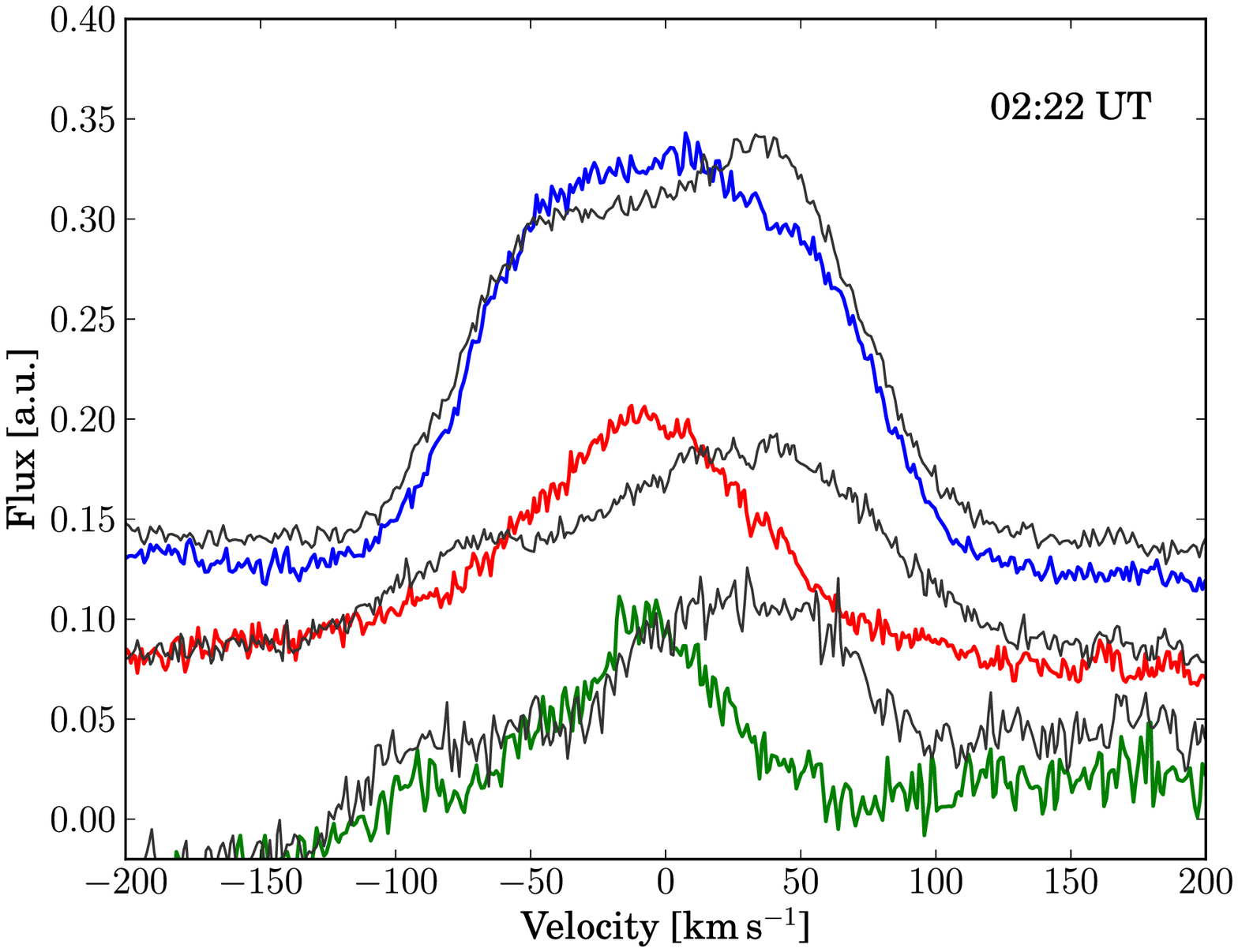} 
\includegraphics[width=5.8cm]{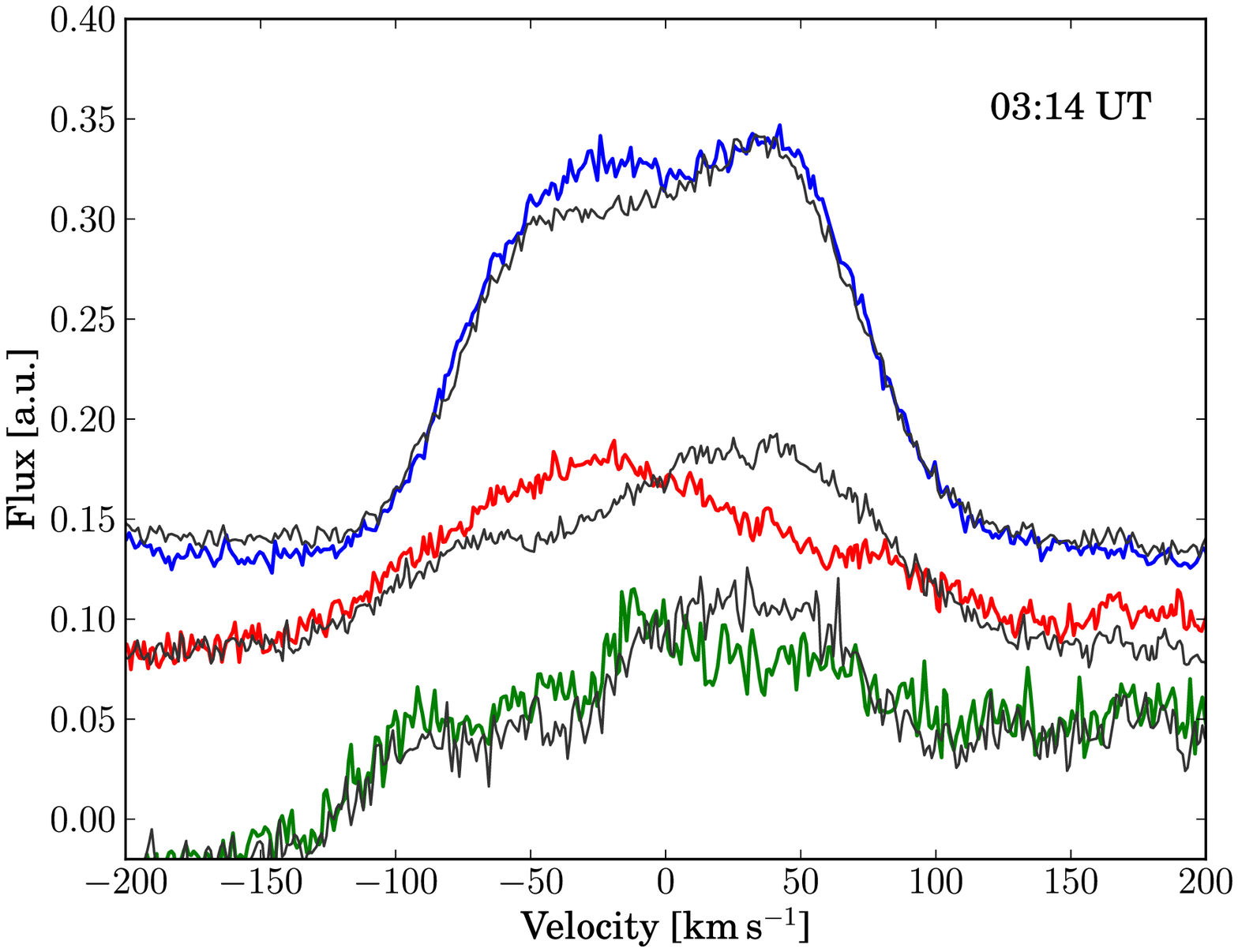} 
\includegraphics[width=5.8cm]{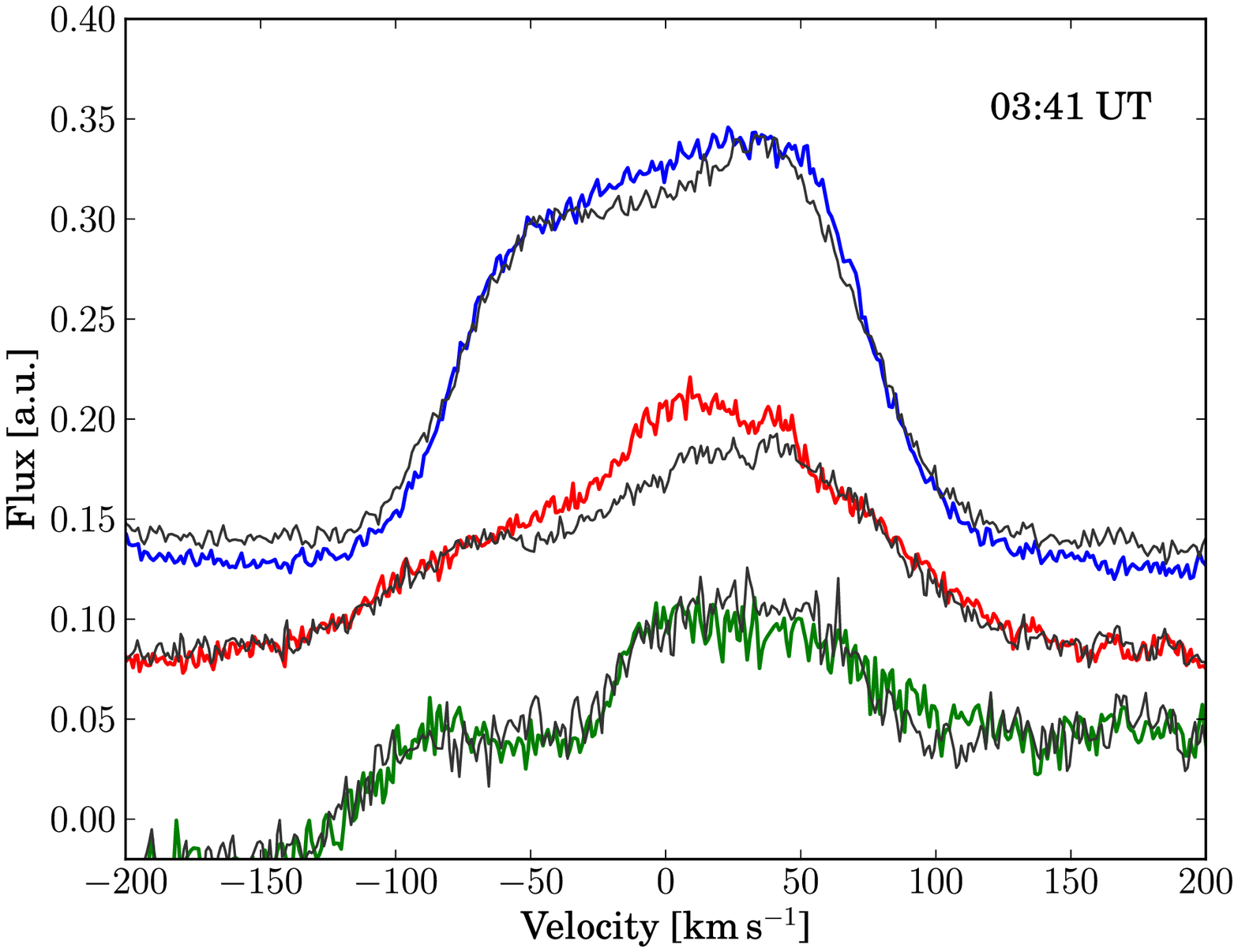} 
\includegraphics[width=5.8cm]{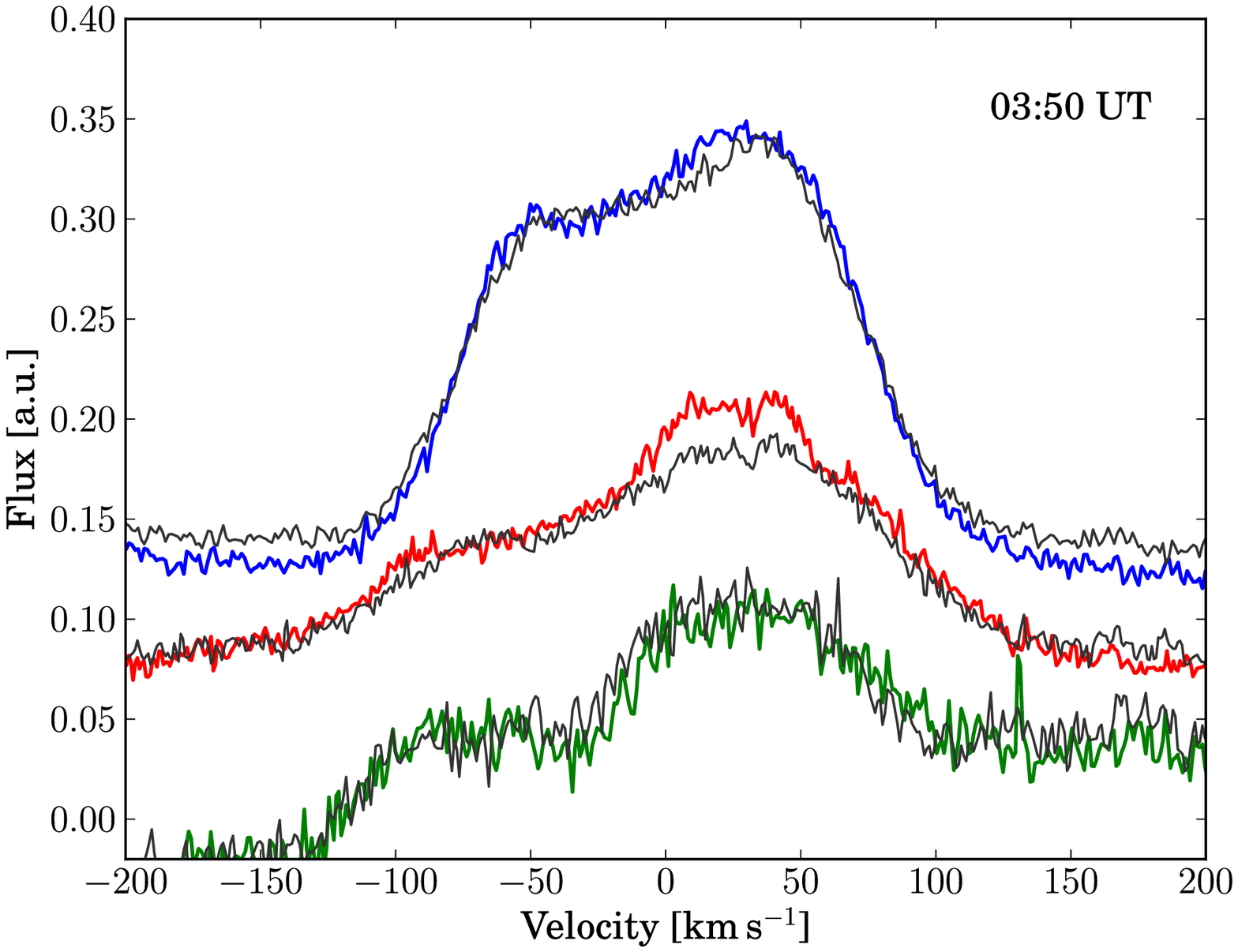} 
\includegraphics[width=5.8cm]{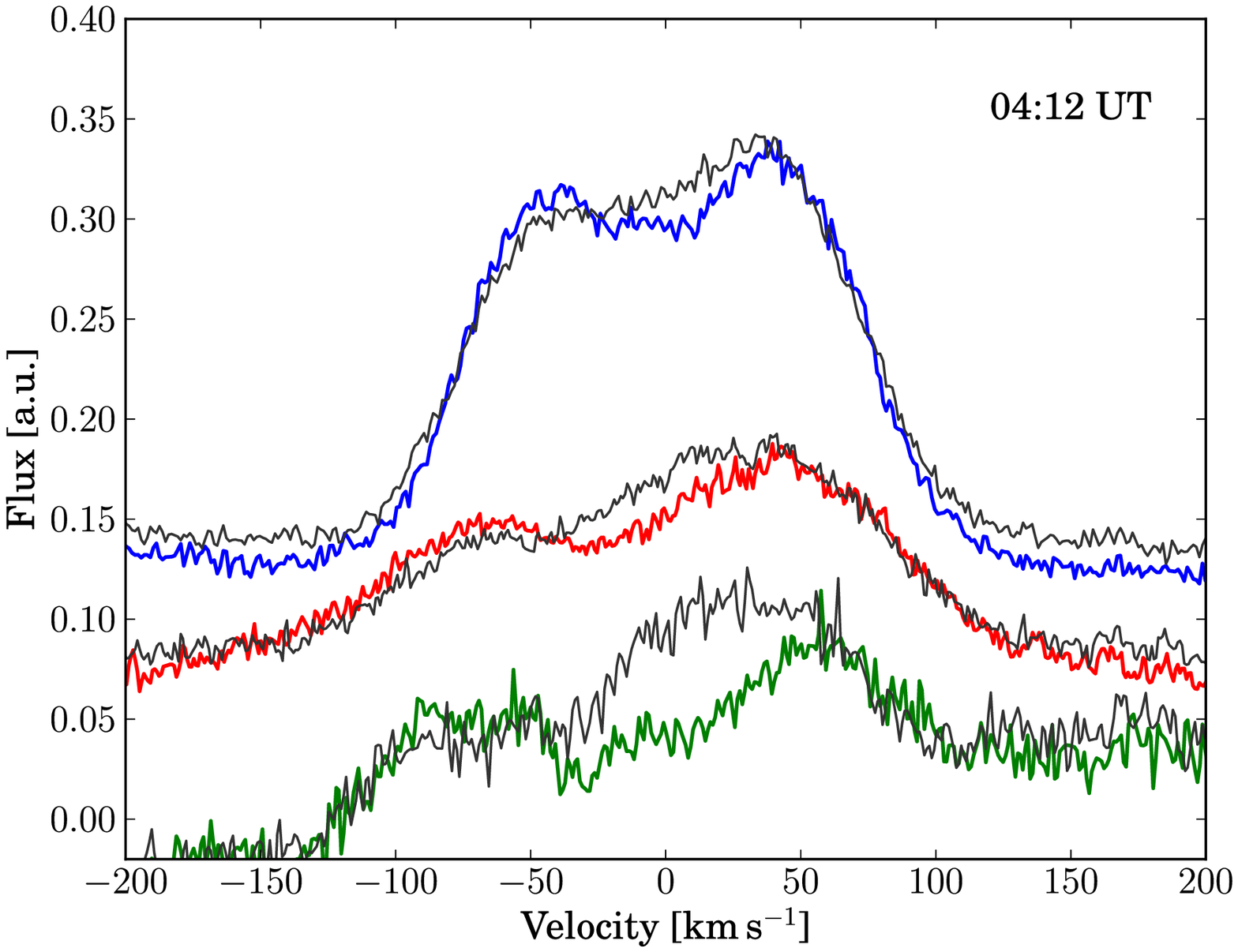} 
\includegraphics[width=5.8cm]{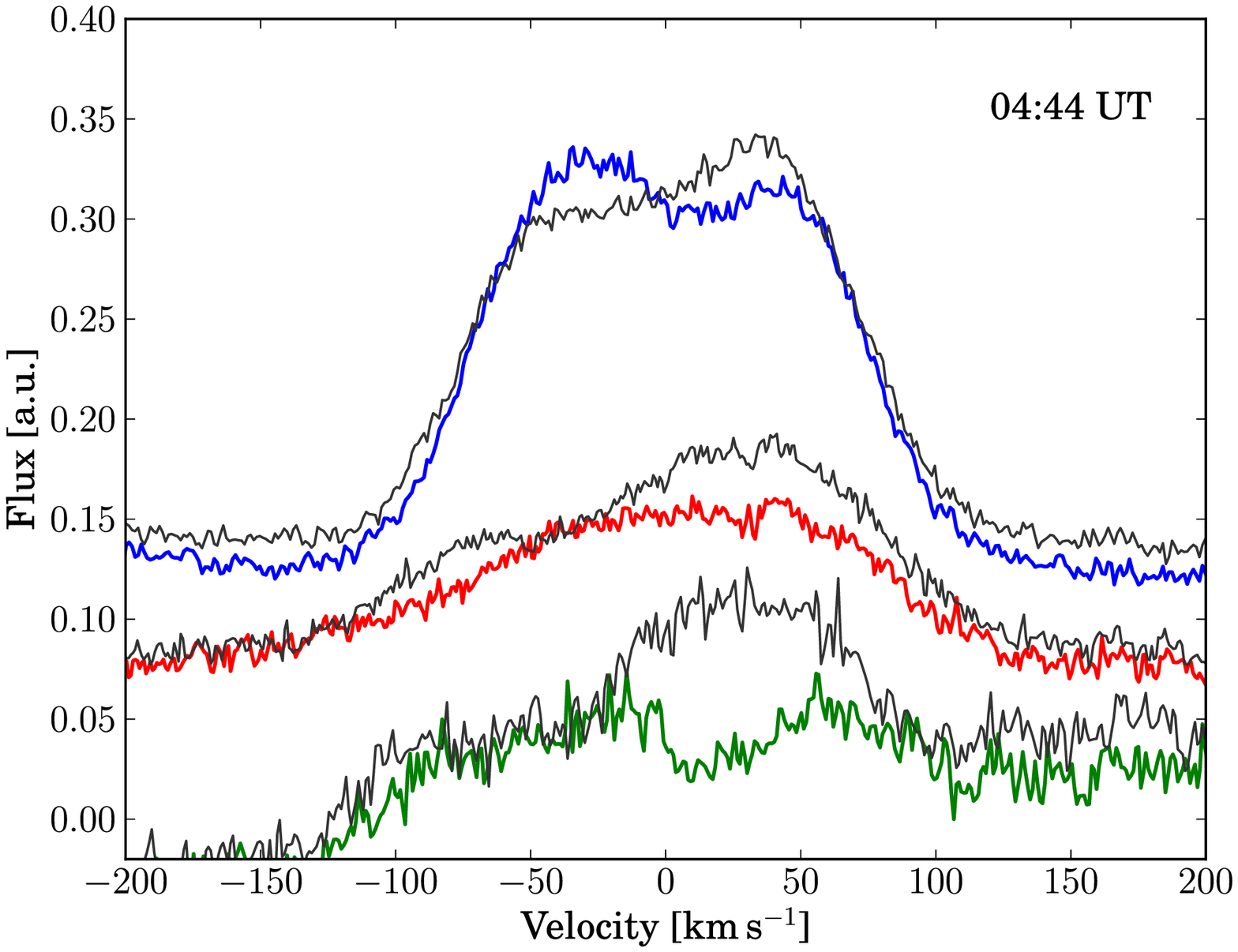} 
\caption{\label{line_comp} Spectra of H$\alpha$ (red), H$\beta$ (green), and  \ion{Ca}{ii} K (blue) 
with a PHOENIX spectrum subtracted and scaled for convenience. 
For comparison the PHOENIX-subtracted flare spectrum (2:58 UT)  
of each line is over-plotted in black.  
Different behaviour is seen for the lines, e.\,g. at 3:14 UT, where the \ion{Ca}{ii} line is still 
double peaked, while in H$\alpha$ and H$\beta$ the peak at 40 km\,s$^{-1}$ has already decayed. Also, the 
rotationally induced shift of line components is seen, e.\,g. in H$\alpha$  
the main component, which moves from about -40 (3:14 UT) to +20 (3:41 UT), +30 (3:50 UT), and finally to 
+50 km\,s$^{-1}$ (4:12 UT).}  
\end{center} 
\end{figure*} 
 
A time series comparing the cores of the three 
strongest lines with a PHOENIX spectrum subtracted is shown in Fig. \ref{line_comp}, where the  
lines are scaled for convenience.  
The lines in the spectrum taken at 2:22 UT show a single-peaked line centre, that 
decays and develops a double-peaked structure \textit{before the flare} for all three lines 
with one peak at about 40 to 45 km\,s$^{-1}$ and a second  broader and  shallower  
peak at about -80 km\,s$^{-1}$ 
for the H$\alpha$ and H$\beta$ lines and about -50 km\,s$^{-1}$ for the \ion{Ca}{ii} K line. This large 
difference in velocity cannot be explained only by different line-forming heights and remains unexplained. 
This second peak can be seen at 6561 \AA~in Fig. \ref{halpha_phoenix} for the H$\alpha$ line in  
the green spectrum taken at 2:48 UT. Also it shows up 
in the flare spectrum in Fig. \ref{line_comp}, since the blue peak is nearly undisturbed by the flare.   
The blue feature is migrating red-wards until it reaches 40-50 km\,s$^{-1}$ at about 4:10 UT  
(see Figs. \ref{line_comp} 
or~\ref{intensitymap}). We interpret the blue peak as the signature of an active region rotating into view.

Event 1 occurred in the red-shifted half of the line, while the blue emission peak 
is nearly undisturbed. After Event 1, for H$\alpha$ and H$\beta$, the line flux  
in the flaring component decays rapidly below the flare onset amplitude, 
while for the \ion{Ca}{ii} K line, the amplitude stays at the level of flare onset 
(see Fig. \ref{line_comp}, 3:14 UT). This indicates that  the heating of the flare site diminishes 
in the upper 
chromosphere, while in the lower chromosphere 
the heating persists. This state persists for about 40 minutes during which the 
blue-shifted component brightens considerably and shifts red-ward with a drift 
velocity of about 90 km\,s$^{-1}$ per hour  
and eventually reaches the about 50 km\,s$^{-1}$ of the original flare site at 
the stage it also reaches maximum brightness (about 3:47 UT). In Fig. \ref{line_comp} 
this corresponds to the peaks seen in H$\alpha$ and \ion{Ca}{ii} K starting at flare  
peak at about -70 (-50 for \ion{Ca}{ii} K) km\,s$^{-1}$, seen at 3:14 UT at about -40 km\,s$^{-1}$, 
at 3:41 UT (and 3:50 UT) the peak reaches about 40 km\,s$^{-1}$. This drift is seen better in the shift of the fitted Gauss kernels for the lines. 
This brightest episode in H$\alpha$ at about 3:45 UT can also be noted in the intensity map 
and coincides with the onset of Event 2.  
Furthermore, another active region rotates into view in the blue line wing, which can 
be seen in the spectrum taken at 4:12 UT (Fig. \ref{line_comp}). The difference between H$\alpha$ and H$\beta$ 
(the H$\alpha$ amplitude is similar to the one during the flare, while the amplitude 
of H$\beta$ is well below the flare values) might be explained by a different  
optical depth of the two lines. While both components 
decay further for H$\alpha$ and H$\beta$, the blue component of the \ion{Ca}{ii} K line brightens until about 
the spectrum taken at 4:45 UT, which again indicates the different heating at different heights.

In the following we discuss the behaviour of the lines during Event 1 in more detail. 
 
\begin{figure}[!ht] 
\begin{center} 
\includegraphics[width=9cm]{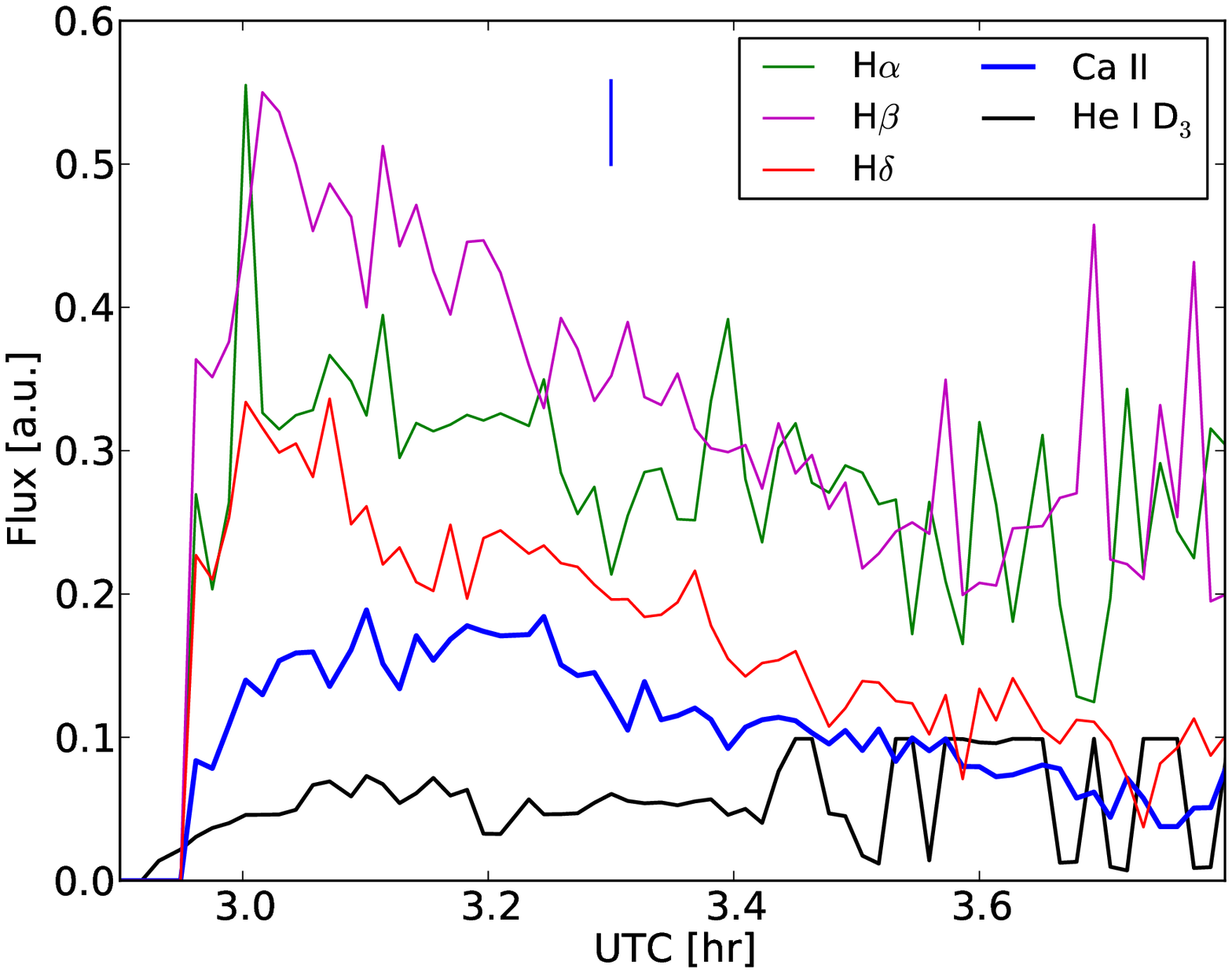} 
\includegraphics[width=9cm]{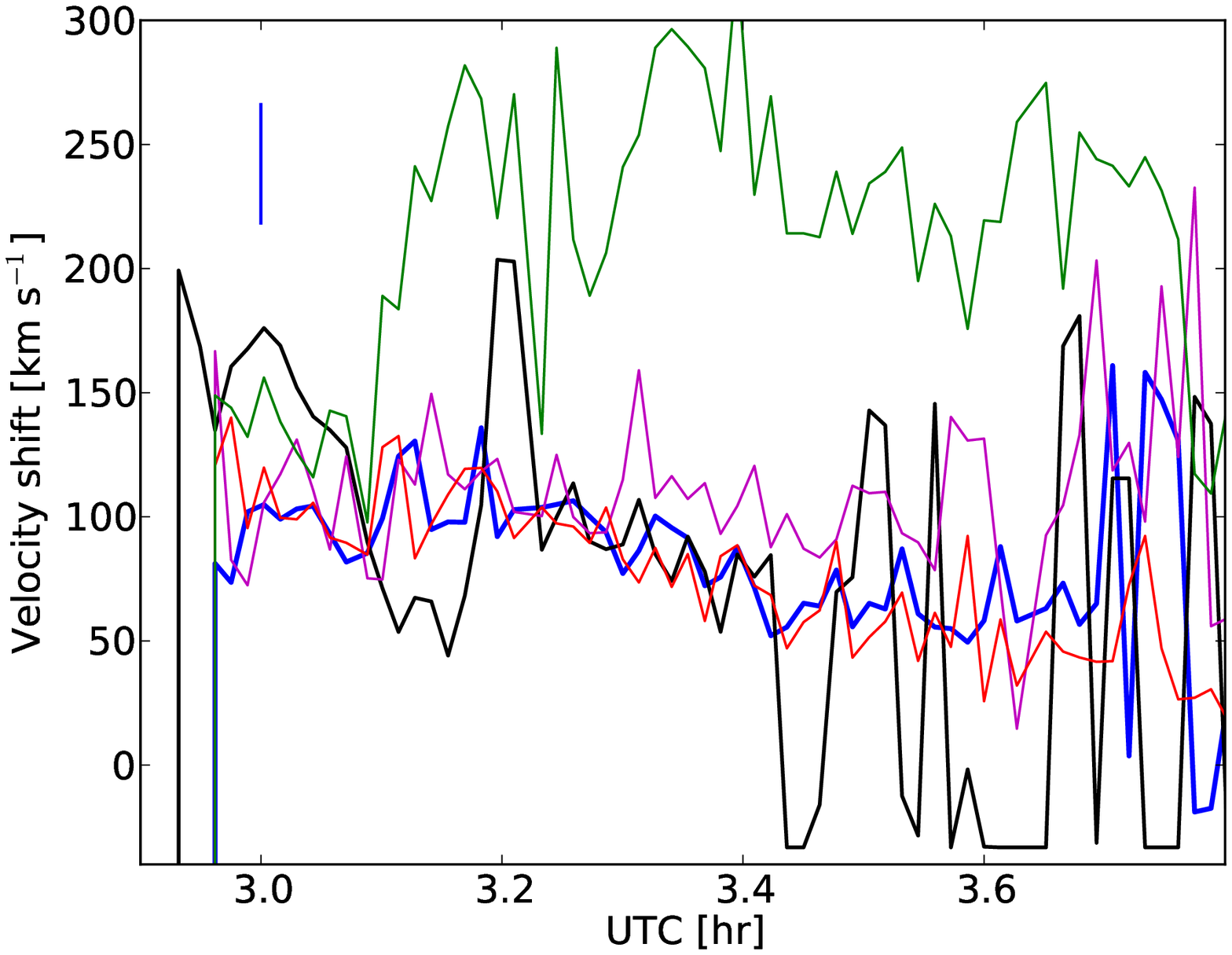} 
\includegraphics[width=9cm]{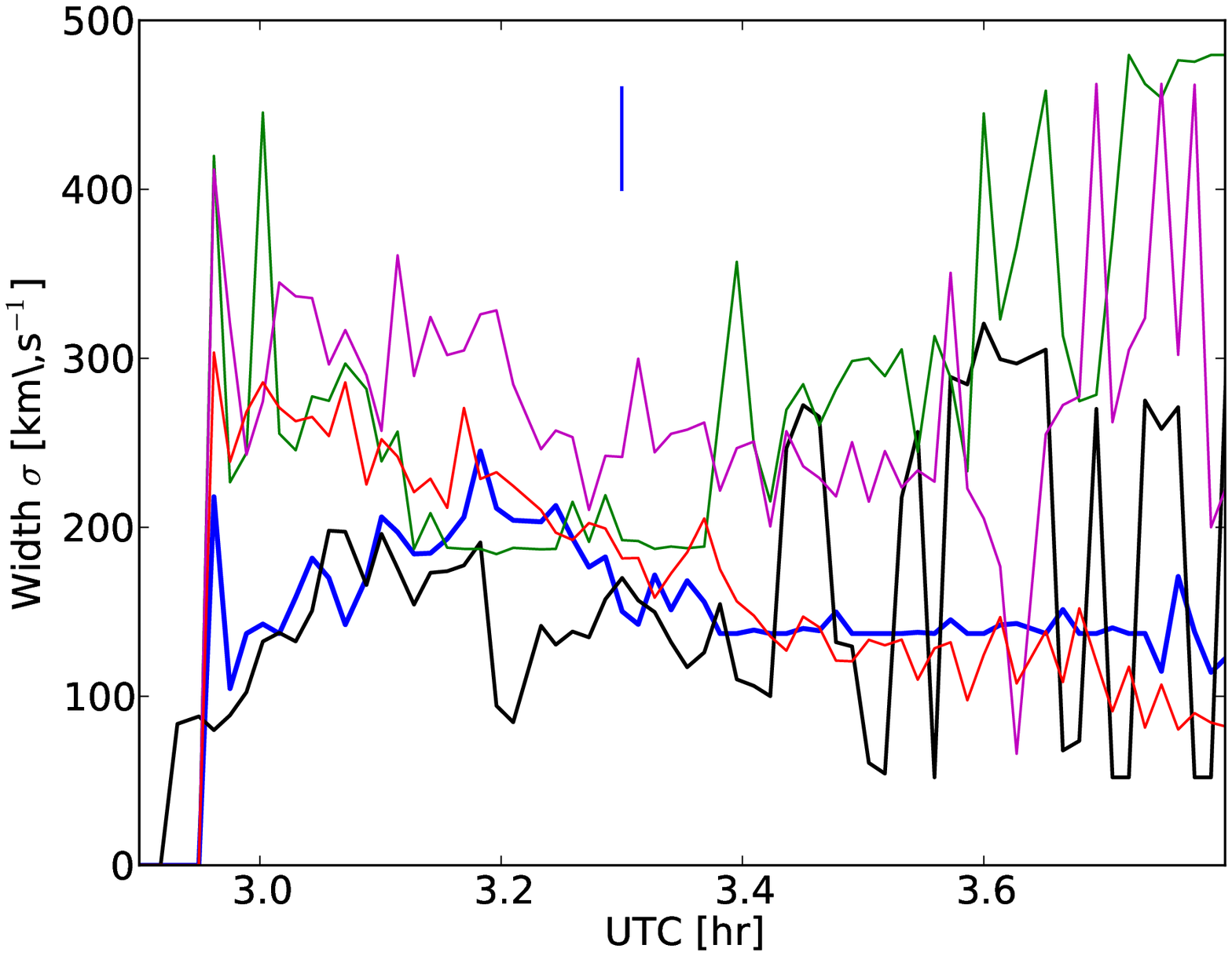} 
\caption{\label{broadcomponent} Characteristics of the broad line component: 
Top: Flux amplitude; Middle:  Velocity shift of the line centre;  Bottom: Gaussian 
width in km\,s$^{-1}$. The legend applies to all panels. 
The earlier rise in the \ion{He}{i} D$_{3}$ line is caused by using a running 
mean for the fitting process. The high velocity shifts in H$\alpha$ are caused 
by line subcomponents influencing the fit. The vertical bar in each panel represents a typical error bar.} 
\end{center} 
\end{figure}

\subsubsection{Broad line components} 
 
During Event 1 the Balmer lines, \ion{He}{i} D$_{3}$, as well as the \ion{Ca}{ii} H and K lines, 
show a very broad component, whose velocities and lifetimes differ significantly from the narrow components. 
For the \ion{He}{i} lines at 3819.6, 4026.2, and 6678.1 \AA\, 
the line amplitudes are not strong enough to reveal a broad component, although the latter 
at least shows a second component. These broad component has been noted mostly in 
M-dwarf flares \citep{Crespo, CNLeoflare, Proxcen}, but also for young K-dwarfs  
(\citealt{Montes} and references therein). They can be 
blue- or (more often) red-shifted, and are often ascribed to a moving turbulent  
plasma component. Broadening, especially of hydrogen lines, may also be caused by 
the Stark effect; for a flare on Barnard's star, \citet{Paulson} attributed 
a symmetric line broadening to the Stark effect. Also, for the Sun \citet{Johns-Krull} 
found evidence of Stark broadening in higher order Balmer lines during a strong flare. 
 
In Fig. \ref{broadcomponent} we show the results of our Gaussian fit to the broad component of 
the lines discussed above. Since individual errors would clutter up the graphs, we only give typical error 
bars for our measurements.   
All broad line components exhibit  
red shifts and no blue shifts. The higher shift 
of the H$\alpha$ line that starts a few minutes after flare onset is most 
probably caused by additional red-shifted narrow line components, which 
are not included in our model but influence the fit of the broad component 
(see Fig. \ref{halpha_diffspec}).

The fluxes of the \ion{Ca}{ii} K  and the \ion{He}{i} D$_{3}$  
line  peak later and more gradually  than the Balmer lines. 
For the  
\ion{Ca}{ii} K line the observed behaviour could be explained by a height effect 
and corresponds directly to the behaviour of the main component (see below). 
For the \ion{He}{i} D$_{3}$ line the broad component is shallow and therefore 
noisy, which may hide a true peak directly after the flare onset. The earlier flare onset for this line is an artefact of 
the running mean used for the fitting and not of physical origin. The broad component vanishes about an hour after 
flare onset as illustrated  by the decaying line flux seen in Fig. \ref{broadcomponent}.

\subsubsection{Stark broadening versus turbulent broadening} 
 
Although the uncertainties of the fit parameters for a our Gaussian line components are rather large
-- as indicated by the scatter of the measurements -- we can discern Stark broadening from turbulent 
broadening in our data. As seen in Fig. \ref{broadcomponent}, there is a clear trend towards Balmer line widths being larger than the widths of 
the helium and \ion{Ca}{ii} K lines from flare onset until about 3:10 UT.
We ascribe this difference to Stark broadening. For a notable Stark broadening of helium lines, 
very high densities of at least 10$^{16}$ cm$^{-3}$ would have to be reached \citep{Chaouacha}, 
which is not expected for this rather small flare. Therefore we argue that the broadening observed in the  
\ion{He}{i} D$_{3}$ and  \ion{Ca}{ii} K is turbulent broadening, while the additional 
broadening observed in the Balmer lines is caused by Stark broadening.  
 
\begin{figure*}[!ht] 
\begin{center} 
\includegraphics[width=18cm]{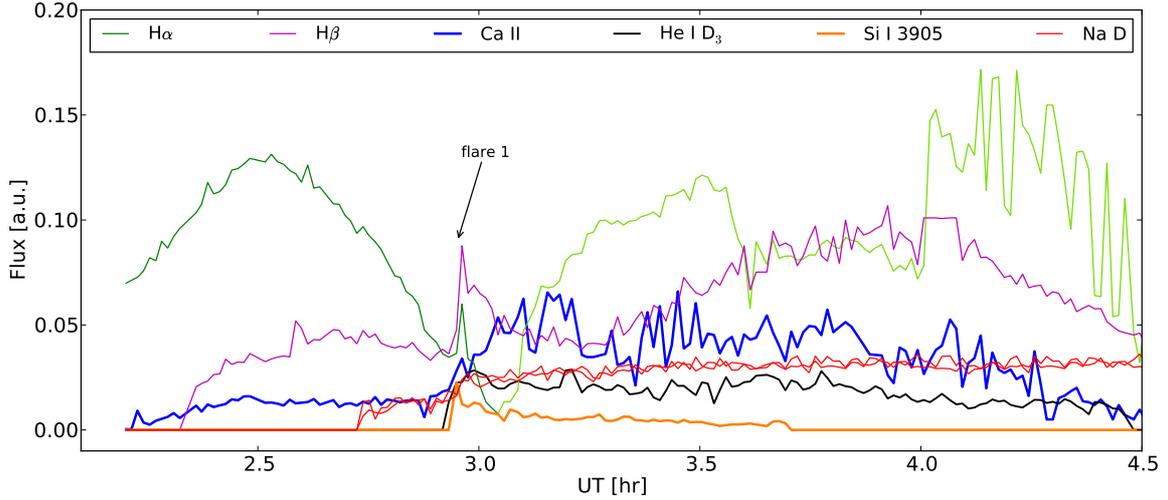} 
\caption{\label{mainflux} Amplitude of the Gaussian fit for the strongest chromospheric emission lines. 
In case of multiple line components,  the strongest narrow fitting component is shown (H$\alpha$, H$\beta$,  
\ion{Ca}{ii} K). The time interval  covers the first 2.3 hours of observations, including the 
flare onset at 2:58 UT. For fitting H$\alpha$, 
a PHOENIX spectrum has been subtracted; the change of the green hue indicates that the fit switched to another line 
component (see text). Colours are indicated in the legend; both \ion{Na}{i} D 
lines are denoted in red.   
} 
\end{center} 
\end{figure*} 
 
We note that Stark 
broadening should affect higher members of the Balmer  series more strongly \citep{Svestka,Worden}.  
While H$\beta$ is as broad as or even broader than H$\alpha$, this is not found 
for the H$\delta$ line. However, the S/N of the spectra 
is decreasing towards the blue as is the amplitude of the Balmer lines with increasing order. 
Therefore, even broader tails may be hidden in the noise. For even higher members of the 
Balmer series, it is not clear whether their broad line components start to merge.   
Moreover, \citet{Paulson} also found less broadening in higher Balmer lines and  
gave NLTE effects as a possible reason, which affect low order lines most.

Although Stark broadening is present in the Balmer lines, we choose to fit 
 the lines with two or three Gaussian components, instead of a Voigt profile. 
 Asymmetric Stark profiles are expected for 
moving plasmas, but would have to be calculated with a full 3D hydrodynamic 
and radiative transfer code, as  done by \citet{Allred2} but is beyond the scope of this work. To obtain 
estimates of the velocities of the plasma movements a Gaussian approximation 
should be sufficient. 

\subsubsection{Narrow line component} 
  
The evolution of the flux of the main narrow line emission component is shown 
in Fig. \ref{mainflux}. The H$\alpha$ line is fitted after subtracting a PHOENIX spectrum  
 and scaled by 0.43, while all other lines were fitted using a quiescent spectrum of AB~Dor~A 
for subtraction. Also, since the H$\alpha$ line displays more than one narrow component, 
the fit switches to another component directly after flare onset, which is marked 
in Fig.~\ref{mainflux}. Also, the dip between 3.5 and 4.0 UT 
in H$\alpha$ flux is not real but is caused by the fit shuffling 
flux from one to another fit component (the total H$\alpha$ flux peaks at about 3.75 UT).

\subsubsection{Subcomponents of the H$\alpha$ line} 
 
\begin{figure*}[!ht] 
\begin{center} 
\includegraphics[width=5.7cm,height=4.7cm]{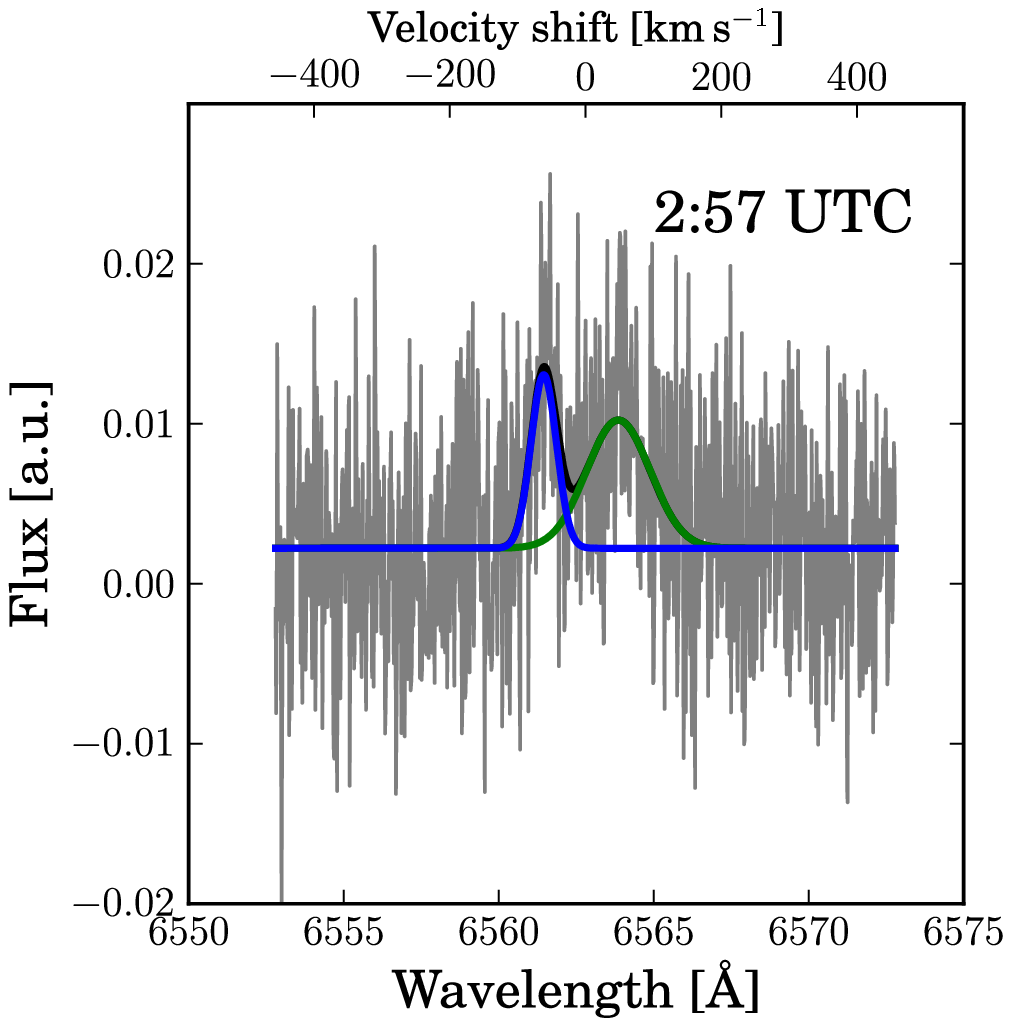} 
\includegraphics[width=5.7cm,height=4.7cm]{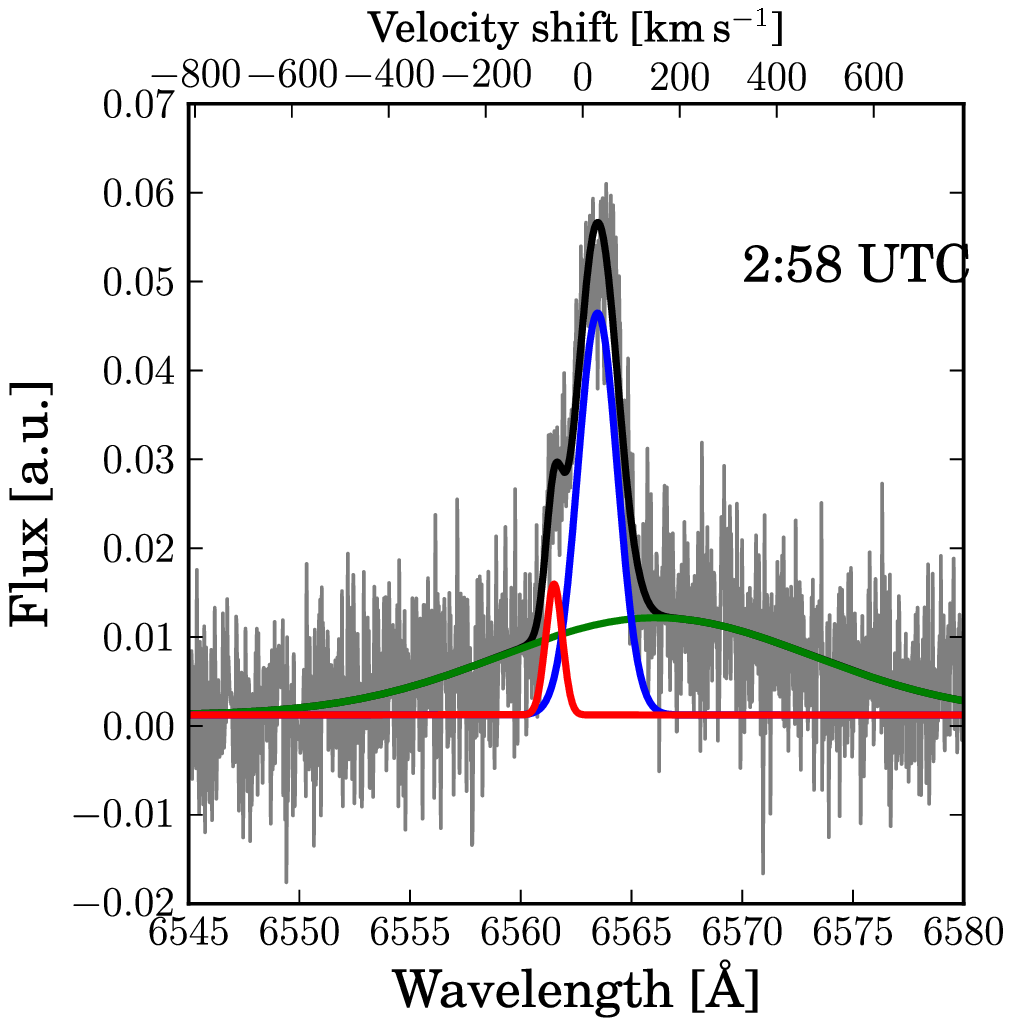} 
\includegraphics[width=5.7cm,height=4.7cm]{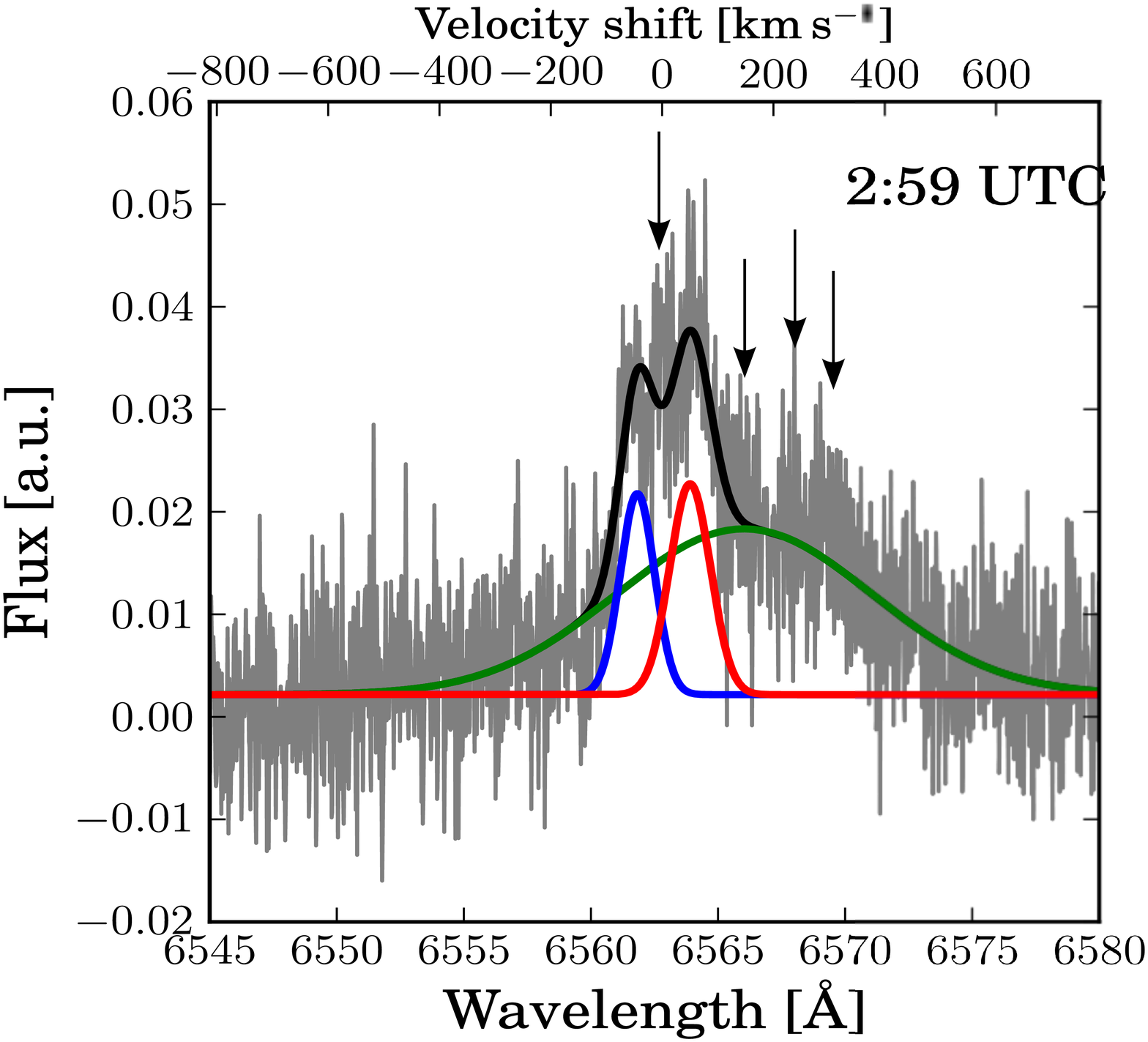} 
\includegraphics[width=5.7cm,height=4.7cm]{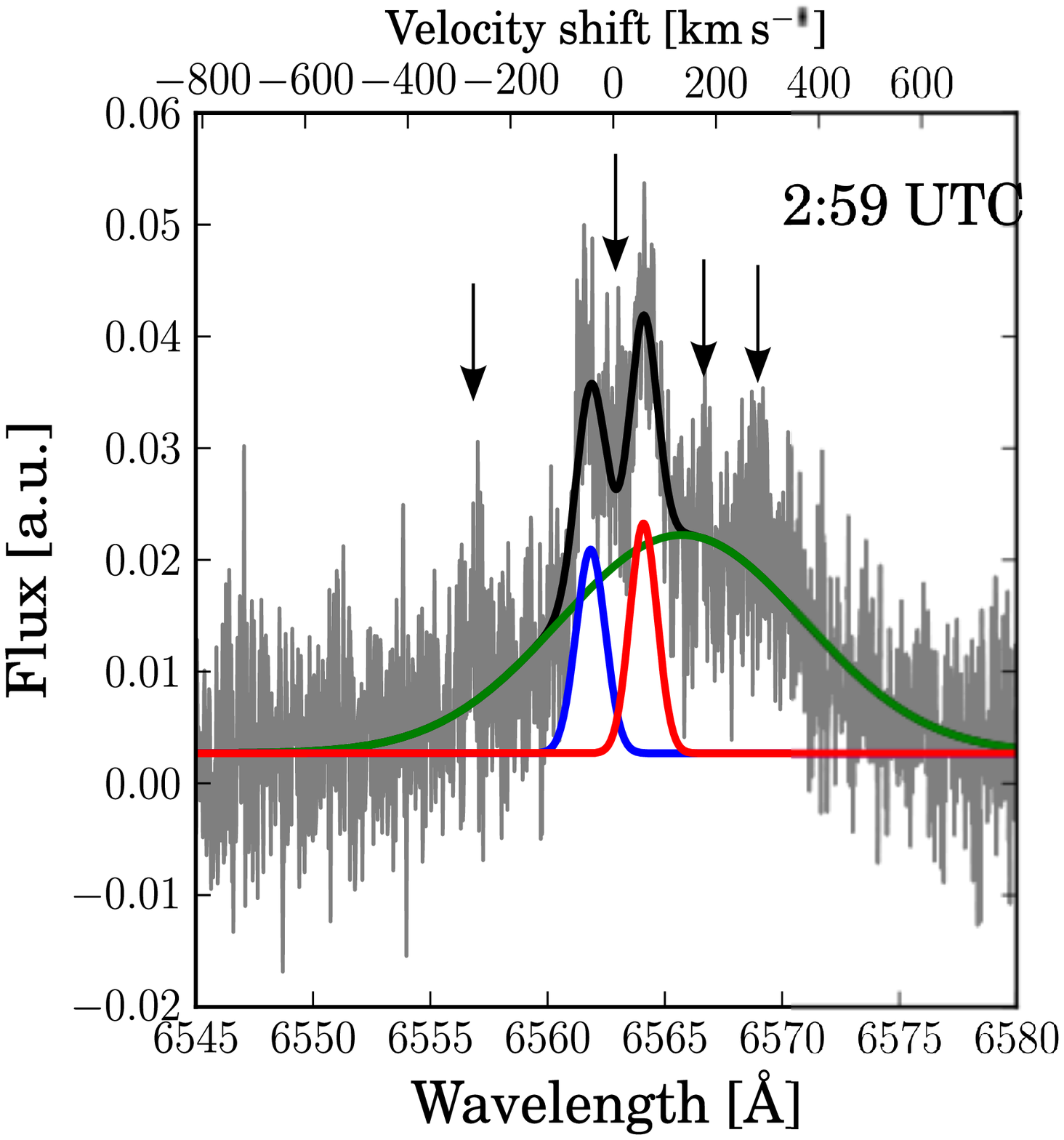} 
\includegraphics[width=5.7cm,height=4.7cm]{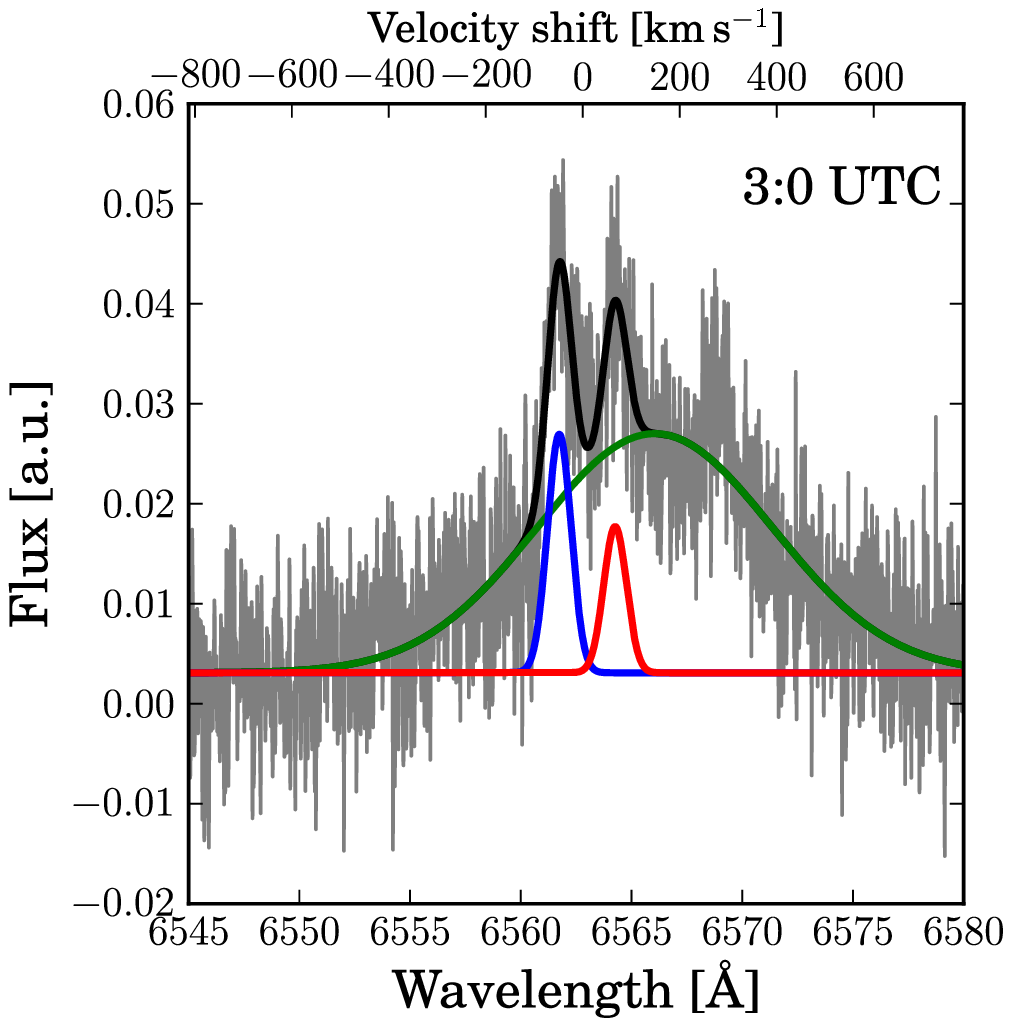} 
\includegraphics[width=5.7cm,height=4.7cm]{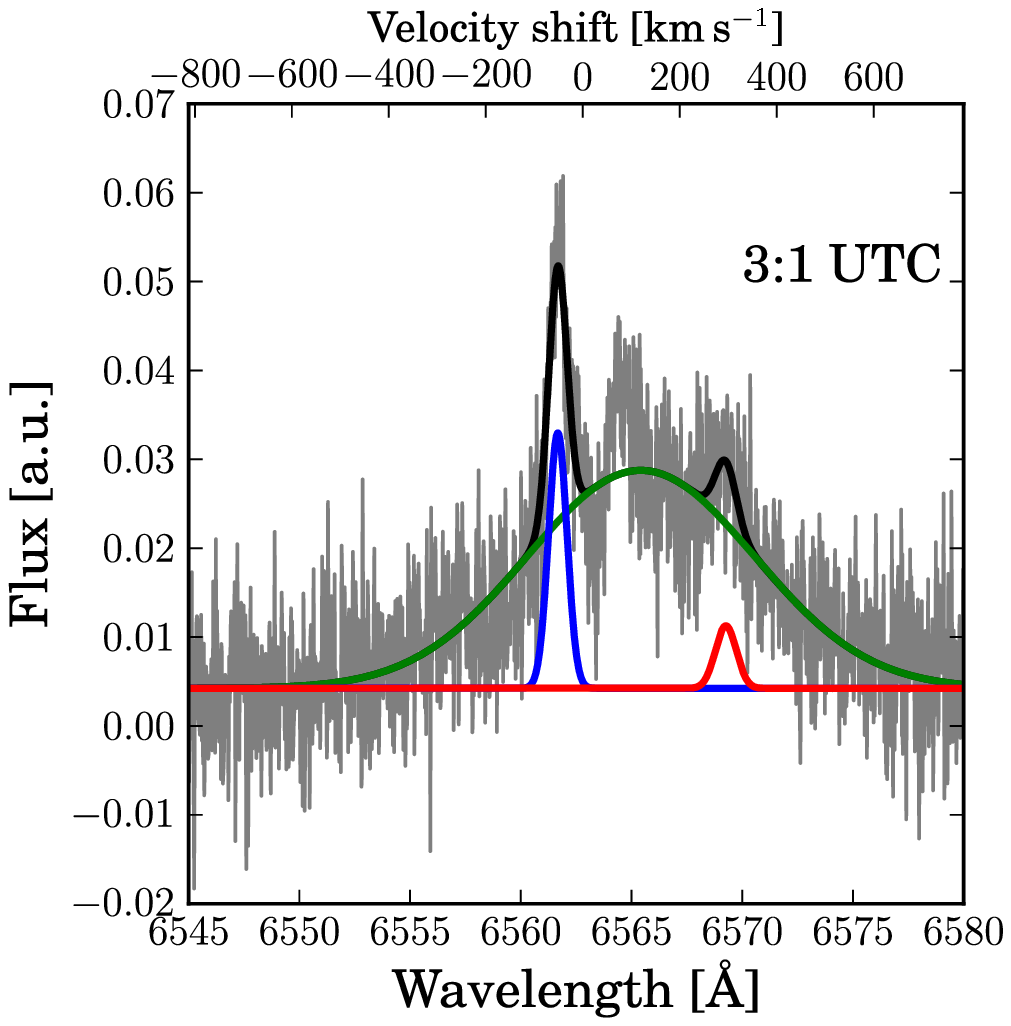} 
\includegraphics[width=5.7cm,height=4.7cm]{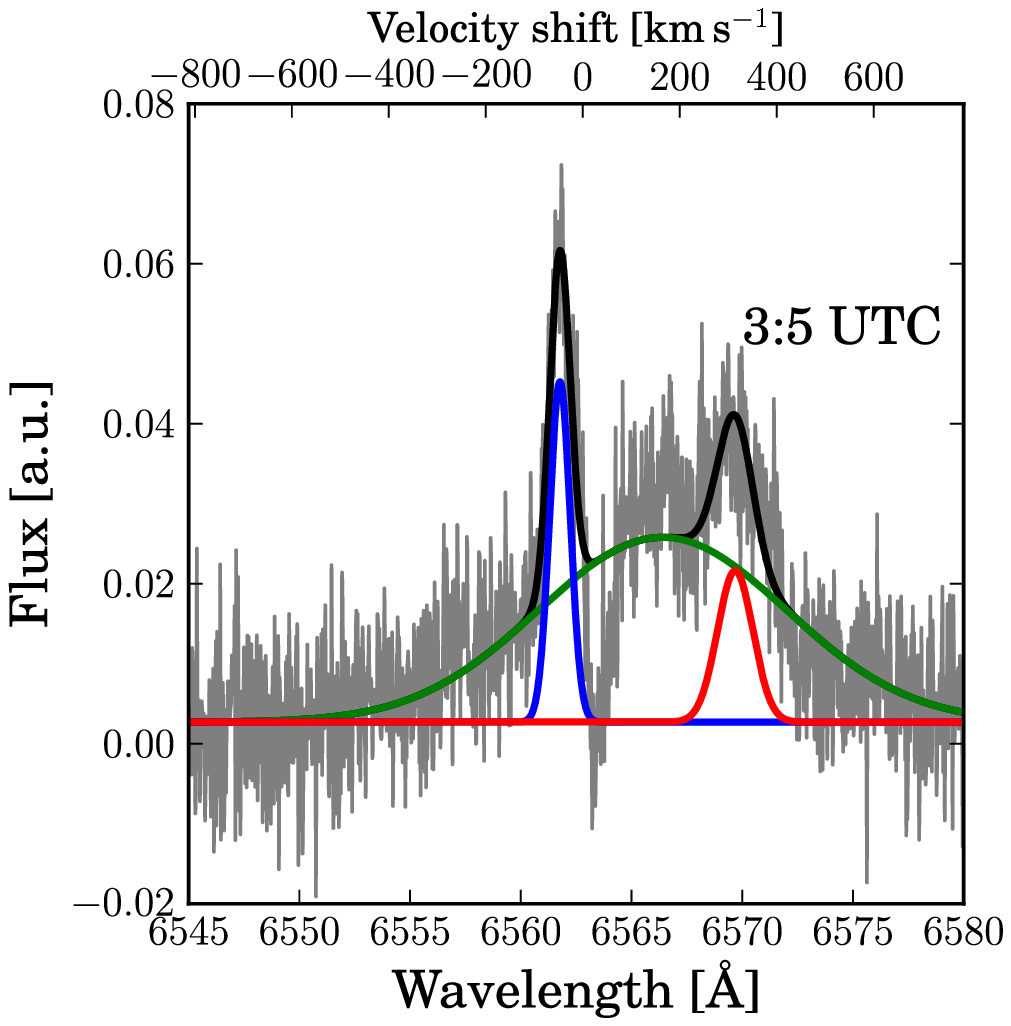} 
\includegraphics[width=5.7cm,height=4.7cm]{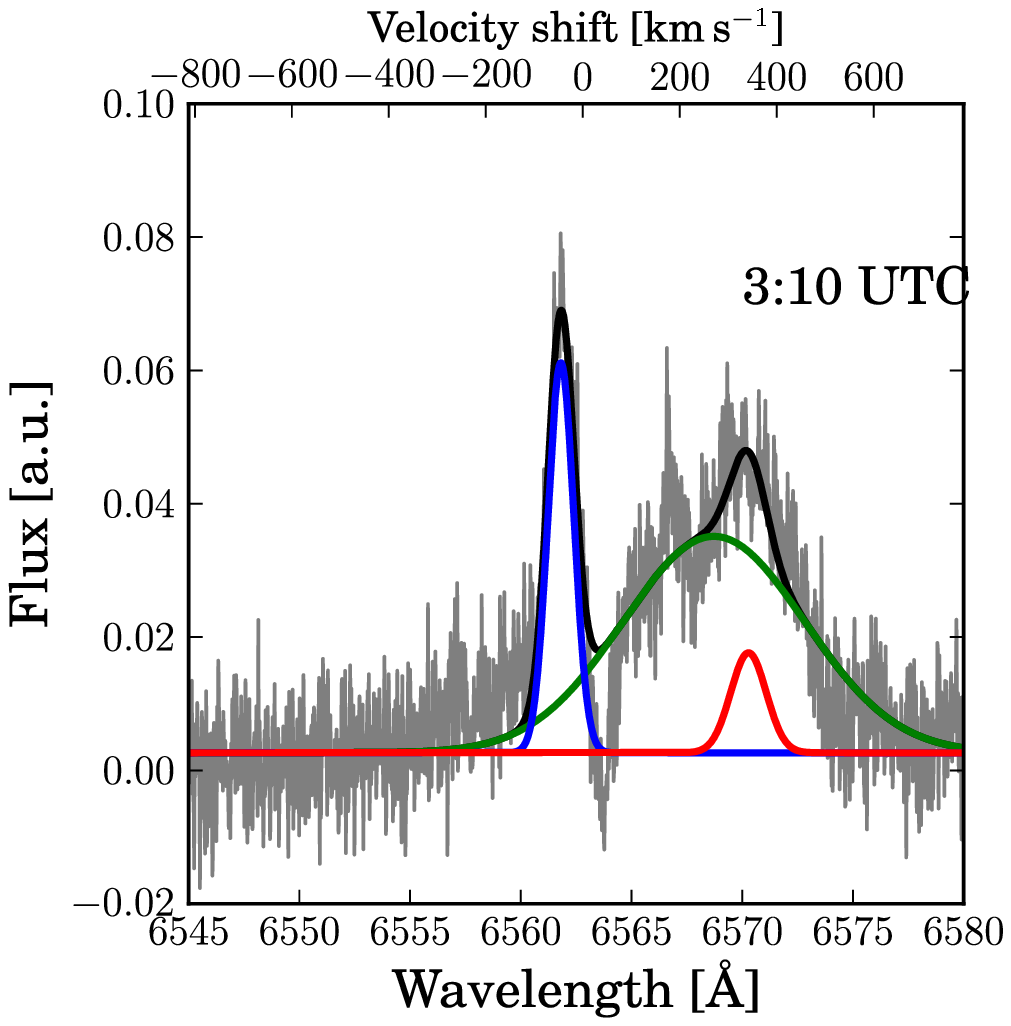} 
\caption{\label{halpha_diffspec}Fit examples of the H$\alpha$ line  
with the observed quiescent spectrum subtracted. The data is shown in grey, the fit 
components are shown in green, blue, and red, and the resulting fit is shown in black.  
The data contains additional narrow short-lived line components, which are not fitted here, but are indicated by arrows for 
the two most complicated spectra at 3:59 UT.  
Between 3:00 and 3:01 UT the fit switches to another line component.  
The second spectrum in the upper row shows the flare onset.  
In the last two spectra one 
can note an absorption line at about 6563 \AA, which is caused by the dimming of a preflare 
line component and not caused by the flare.}
\end{center} 
\end{figure*} 
 
We also searched for weaker subcomponents in 
the H$\alpha$ line. As discussed in Sect. \ref{obsvssim} they are more pronounced, when 
subtracting a quiescent spectrum of the star instead of a PHOENIX spectrum.  
Figure \ref{halpha_diffspec} shows the evolution of the H$\alpha$ line \textit{only during a few minutes} around flare onset; 
for all panels, the averaged spectrum taken between 2:56 and 2:58 UT has been subtracted as a  
proxy for the quiescent state. The figure nicely illustrates the complex and rapidly evolving multi-component structure of the 
chromospheric emission  during the impulsive phase of the flare.  
Obviously, for most spectra a fit with only three components is not sufficient to cover all 
components. However, a fit with four or more components is even more 
unstable than a fit with three components and can usually not be constrained by the data.  
In addition to the main flaring component, there 
are several short-lived and a few longer lived narrow components  of which the 
bluest at about 6561 \AA\, is growing stronger during the decay of the flare. At 3:05 UT, about ten  
minutes after flare onset, an absorption component appears to emerge. However, it is most likely 
not caused by the flare, but results from changes in the non-flaring 
line components relative to the preflare state.  
On the other hand, we argue that the subcomponents in the H$\alpha$ line, appearing during 
the first few minutes of the large flare, are real, since notable changes in the non-flaring spectrum seem to appear on 
time scales of about ten minutes, while during the flare the development of line features
(appearing, shifting, and disappearing) takes place on much 
shorter time scales.  

We interpret these weak subcomponents of the H$\alpha$ line  as signatures of shock waves in the chromosphere.  
Alternatively they could be  small dense `blobs' created during the flare, moving 
at high intrinsic velocities. Some of them can only be  
identified in one or two consecutive spectra, translating into a lifetime of one to two minutes. We counted 
up to six such subcomponents in a single spectrum (2:59 UT, two spectra after flare onset), moving with velocities ranging from 
-270 km\,s$^{-1}$ to +260 km\,s$^{-1}$. If these subcomponents are present in more than one spectrum, they 
normally decelerate; i.\,e., blue-shifted features shift red wards, while red-shifted features show a drift to 
bluer wavelengths. This strengthens our confidence that these features are not artefacts and are really 
caused by the flaring material. 
The only other line besides H$\alpha$ with a strong enoughS/N level to identify subcomponents 
is the \ion{Ca}{ii} K line. There we find a comparable number of subcomponents, but we cannot  
identify any of the \ion{Ca}{ii} subcomponents with H$\alpha$ subcomponents even for the stronger ones.  
This is caused partly by  
the different velocity scatter ranging from -390 to +190 km\,s$^{-1}$. Actually, this is 
 not surprising, if these subcomponents are indeed tracers of material of different temperature 
potentially at different heights of the atmosphere.


\section{Discussion}\label{discussion} 
 
\subsection{Comparison between X-ray and optical signatures}\label{X-ray-optical} 
 
During our 58\,ks exposure of XMM-Newton three medium-sized flares occurred within 
a time interval of only six hours, while the remaining light curve showed no major 
flux enhancement. In the same time interval the OM light curve also 
shows larger flux variations, coinciding with the X-ray events. The chromospheric emission lines 
show more complex flux variations, but it is certainly correct to assume that 
the different wavelength bands actually trace the same events in different parts 
of the atmosphere. Furthermore, the question arises whether the three events originate in the same 
loop or arcade. In the following we discuss these issues for each of the three 
events. 
 
\subsubsection{Event 1} 

\begin{figure} 
\begin{center} 
\includegraphics[width=8cm,clip]{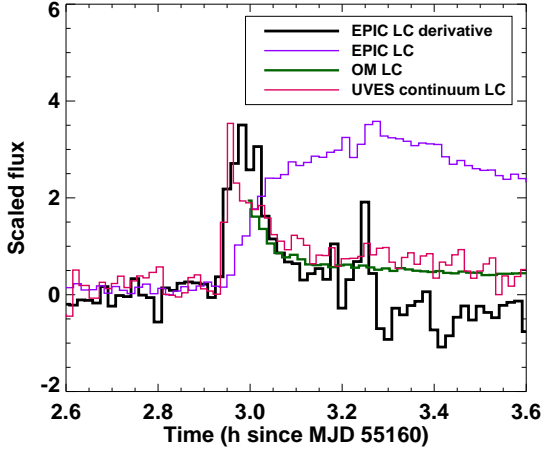}\vspace{-0.4mm} 
\caption{\label{neupert} Neupert effect observed during the large flare on AB~Dor~A. Depicted are the combined 
EPIC X-ray light curve in violet, its time derivative (smoothed by five bins) in black, the OM light curve in green, 
and the UVES continuum light curve in red.} 
\end{center} 
\end{figure}

Comparing the X-ray and optical/UV light curves around the main flare 
onset, we found strong evidence of the Neupert effect.
The chromospheric emission lines also reacted and peaked at 2:58 UT, 
while the X-ray emission peaked later at 3:16 UT. 
The time integral emission due to accelerated particles 
(like $H\alpha$ emission, white-light emission) 
resembled the rise of the flare light curve in  
the soft X-rays, a phenomenon denoted as the Neupert effect. 
In Fig. \ref{neupert}, we plot the time derivative of the soft X-ray light 
curve (using only EPIC data) and the optical light curve. Since there is no 
OM data available during the flare rise we construct an optical  
light curve using the UVES continuum spectra between the wavelengths 3895~\AA ~-~3920~\AA.  
We note that during flare rise the derivative of the X-ray light curve matches the shape of the  
optical light curve; it is also obvious that the optical/UV peak precedes the 
X-ray peak  (see Fig. \ref{neupert}).

The chromospheric lines do not react strongly to the flare in amplitude. In contrast to  
the X-ray and optical/UV light curve, they show even stronger emission before and after the flare 
events; i.\,e., 
the flare does not dominate the chromospheric line emission.  
Despite the weak reaction of the chromospheric lines in amplitude, 
the strong lines do show turbulent broadening, the Balmer lines even show 
Stark broadening and, on top of that, in the H$\alpha$ line profile
single short-lived shock-like events
can be identified. All this is accumulating evidence that 
the above-mentioned picture of a flare affecting different atmospheric 
layers is certainly correct for this event.

\subsubsection{Event 2} 
 
\begin{figure} 
\begin{center} 
\includegraphics[width=9cm]{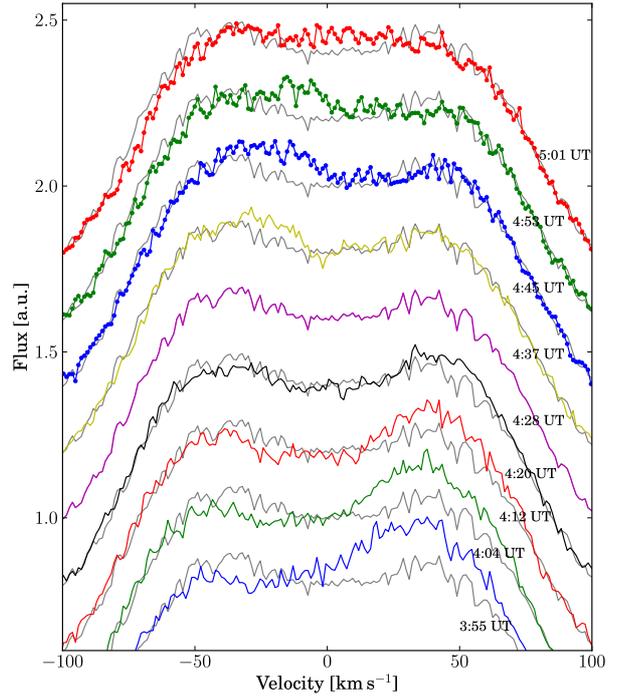} 
\caption{\label{comp_ca3930} Central part of the \ion{Ca}{ii} K emission line for 
different times. The spectra are offset for convenience, and 
a PHOENIX spectrum has been subtracted. The spectrum taken at 4:28 UT is always over-plotted 
for better comparison.} 
\end{center} 
\end{figure}

Event 2 has its onset at 3:40 UT and its broad X-ray maximum between 
about 4:20 and 4:50 UT followed by a slow decay until about 6:00 UT.  
While the onset coincides with the brightest episode in H$\alpha$-emission, 
the broad X-ray maximum has no clear counterpart in the chromospheric lines. 
Between about 4:15 and 4:45 UT, there is a slight brightening 
in one of the H$\alpha$-components at a velocity of about -35 km\,s$^{-1}$, 
see Fig.~\ref{line_comp}.
This brightening is not strong enough to be noticed in the intensity map 
of Fig. \ref{intensitymap} or 
in the main line component analysed in Fig. \ref{mainflux}. 
It is most pronounced in 
the \ion{Ca}{ii} K line, where it even shows up in the intensity map as 
a slight brightening. As an example, we show the central part of the \ion{Ca}{ii} K line 
covering the times of interest in Fig. \ref{comp_ca3930}.  
The feature is not seen in H$\beta$, which may be due to the noise level 
in this line. 
 
 Because of the slow rise and decay of the peaks in the X-ray and the chromospheric 
 light curves, it is hard to decide, whether  
this brightening of the chromospheric lines 
 is physically connected to the X-ray 
brightening. Therefore, 
the X-ray signal may stem from a reheating event, while in the chromosphere, 
a different active region undergoes a brightening that only coincides in time 
with the coronal activity.

\subsubsection{Event 3} 
 
Event 3 is characterised by a peak at 7:46 UT in X-rays, starting at 7:34 
and lasting until 8:16 UT. 
Unfortunately, the third flare partly falls in a gap of the UVES observations.  
Nevertheless, the flare onset is covered and observed at about 7:34 UT, both,  
in the OM and in the chromospheric emission lines.  
The chromospheric emission lines all show a rather weak reaction compared 
to X-rays and the OM. They exhibit a very broad peak 
with the \ion{Na}{i} D  lines showing increasing flux until the end of the  
observations at about 9:20 UT. 
\ion{Ca}{ii} K peaks at about 9:00 UT, while H$\alpha$ appears to 
peak during the observational gap.

\subsection{Comparison to other work and to the radio data} 
 
The timing  of different flares for multi-wavelength observations has been extensively 
discussed by \citet{Osten}, who compare parallel radio, optical, UV, and  
X-ray observations for the young flare star EV~Lac and found that flares in different 
wavelength bands need not have counterparts in any other wavelength bands.     
Also \citet{Kundu} found little correlation between radio and X-ray variations 
on the flare stars UV Cet, EQ Peg, YZ CMi, and AD Leo. While these two studies did not find a 
strong correlation between different wavelength bands, other studies did. 
For example, \citet{Liefke} in their study of the flare star CN Leo found clear
correlations between X-ray emission, chromospheric lines, and the photospheric continuum 
at least for the larger events, while smaller events were not seen in all wavelength bands.

During Events~1 and 3 all line fluxes,  
X-ray, optical, and chromospheric, respond to the flare. On the other hand, 
Event~2 is substantially different: this long-lasting heating event is 
observed mostly in the X-ray light curve with little, if any, associated 
variability in the optical light curve (as seen in the OM) and chromospheric  
line fluxes.
Therefore, Event 2 appears to be largely confined to the corona. 
In agreement with earlier 
studies, we could not find any correlation between the radio 
light curve and other wavelength bands. Furthermore, 
the radio light curve does not show a significant reaction during the  
prominence crossings.

\subsection{Location of the flaring regions}\label{velocity_discrepancy}  
 
\begin{figure} 
\begin{center} 
\includegraphics[width=9cm]{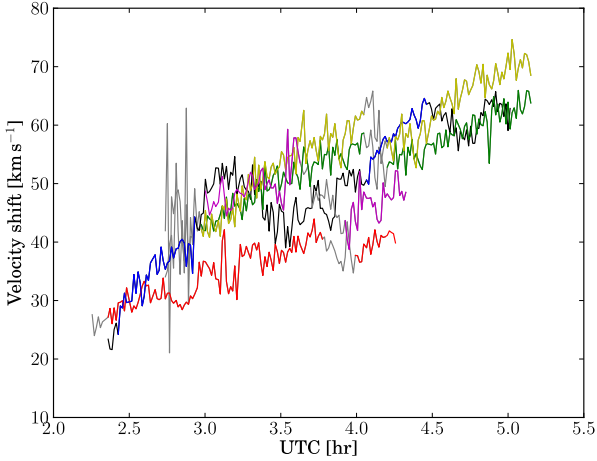} 
\caption{\label{driftvelocity} 
Velocity shift of strong emission lines:
\ion{Ca}{ii}~K (main component, red),
H$\beta$ (main component, blue),
\ion{He}{i} D$_{3}$ (magenta), and
\ion{Na}{i}~D (two components, rendered green and yellow).
Grey and black indicate
less reliable sections of the H$\beta$ curve. 
The graphs of \ion{Ca}{ii}~K and \ion{He}{i} D$_{3}$
are truncated after about 4:20 UT because of the large scatter of the measurements.} 
\end{center} 
\end{figure}

During Event 1 the chromospheric lines listed in Table  \ref{linetable} and the main components of the Balmer lines 
show a mean velocity shift of 39.6 $\pm$ 9.6 km\,s$^{-1}$ . 
Most of these lines appear during the flare onset, but the Balmer lines and other strong lines 
can be identified before the flare as well. These strong lines show a slow velocity drift 
over a long time interval, but no 
abrupt change in their velocity at the onset of Event 1  as 
can be seen in Fig. \ref{driftvelocity} for a time interval extending well into flare Event 2. 
This suggests that the projected rotational velocity of the star and not the intrinsic velocity of the flaring material 
dominates the velocity of the main line component. 
This interpretation is further supported by observations of slowly rotating M-dwarfs, for 
which line shifts during flares are normally not observed; for example,
\citet{Reiners} searched for radial velocity jitter 
in UVES observations of the M-dwarf CN Leo and found a jitter of 
below 1 km\,s$^{-1}$ even during the 
observed very large flare. 
Nevertheless, most of the glitches and jumps visible in Fig. \ref{driftvelocity} 
in the \ion{Na}{i}~D, \ion{Ca}{ii}~K, and \ion{He}{i} D$_{3}$~lines occur between the onset
and maximum of Event 2. 
Whether they are physically connected to this event is not clear, though. 
  
Under the assumption that the velocity shift in the chromospheric lines is dominated by rotation and that 
the chromospheric line emission originates in a region very close to the stellar surface, one can  
try to locate the flaring active region. 
Using an inclination of 60$^{\circ}$ of the stellar rotational axis 
\citep{Kuerster1, donati} and 90 km\,s$^{-1}$ as projected rotational velocity, the  active region 
would be about 25$^{\circ}$ off the centre of the stellar disk in longitude, if  
located at the equator, and at about longitude of 60$^{\circ}$, if  located at 
60$^{\circ}$ latitude.  
Thus, the possibility of the flaring active region being circumpolar  
(i.\,e., visible throughout the full rotation cycle) is excluded on the grounds that  
there are no reversals in the observed velocity shifts (see Fig. \ref{driftvelocity}).
However, we note that a small velocity shift 
reversal could 
be masked by the changing components of the 
emission lines during the flare.

To constrain the location of the active region further in latitude, we computed the times when the flare 
would still be visible for different latitudes. However, for the 45-minute flare duration (the  
time span for which the metal lines and the broad line component can be identified), only latitudes 
lower than about -30$^{\circ}$ are excluded. As another approach, we measured the 
drift velocity, 
i.\,e. the slope of the velocity shift in Fig. \ref{driftvelocity}, 
of the lines to be about 20 km\,s$^{-1}$\,h$^{-1}$ during Event 1 and large parts of Event 2. 
This drift velocity implies  a latitude of about 60$^{\circ}$. 
However, the highest possible velocity for this latitude on the surface of the 
star is about 45 km\,s$^{-1}$, 
which is significantly exceeded by the measured velocities reaching about 70~km\,s$^{-1}$,  
see Fig. \ref{driftvelocity}.
A possible solution would be to locate the flaring region at some distance 
from the stellar surface, but distances up to one stellar radius also do not give consistent results, 
so there seems to be some inconsistency in the drift velocity, which may be influenced to some extent by the line profiles. 
Thus in summary, we cannot consistently locate the active region 
responsible for Event 1.  
 
Although the physical connection between the chromospheric and coronal activity 
is unclear for Event 2, we try to locate the chromospheric active region.  
To complicate the situation even more, the measured velocities of the chromospheric 
 material at the onset (40-50 km\,s$^{-1}$) and the 
maximum (-35 km\,s$^{-1}$) of Event~2 differ substantially, 
suggesting that the emission may not originate in the same active region.  
For Event~3 the measured velocity is~-~24 km\,s$^{-1}$. These velocities, together with 
the rotational drift of the lines, suggest there is no common origin of the different events. This 
would only be possible for a circumpolar active region, a case excluded above. For instance, Event~2  
may even be a superposition of smaller chromospheric brightenings in different regions. 
Also the intensity map in Fig. \ref{intensitymap} suggests different active regions 
for the Events~1 to 3. Therefore, the simple scenario of a reheating or flares in different arcades of 
the same active region  
does not seem to be justified despite the close 
succession of the three events.

\subsection{Geometry of the flaring region}

\begin{figure} 
\begin{center} 
\includegraphics[width=8cm]{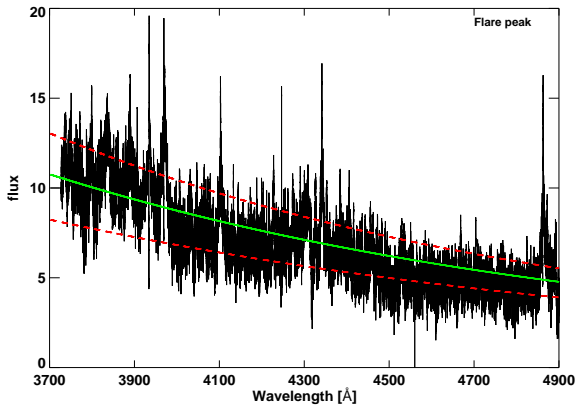} 
\caption{\label{temp_opt} Blue UVES spectrum with the quiescent flux subtracted, covering the peak of flare Event 1.  
Overlaid in green and red are different black-body fits.} 
\end{center} 
\end{figure}

The continuum of the flare optical spectra shows a well-defined slope, suggesting a 
black-body origin.  To compute the black body temperature we use the flare spectrum with
the  quiescent flux subtracted. 
The black body fit gives a temperature estimate of  
16 000 K during the flare peak. Figure~\ref{temp_opt} shows the best-fit blackbodies and   
black body fits with fixed temperatures at 14 500~K and 17 500~K for comparison.  
 
Using the UV light curve obtained with the OM, the optical filling factor can be estimated. 
For the OM we find a mean count rate of 70 cts/s during the quiescent phase between 21:00 UT and 01:00 UT. 
Additionally, we obtain the count rate of 190 cts/s at the flare peak (Event 1). Using the count-to energy conversion  
factor of 1.66$\times$10$^{-15}$ erg/cm$^2$/s for a star of spectral type K0 from the \emph{XMM-Newton} handbook,  
we calculated a flux of  
1.1$\times$10$^{-13}$ erg/cm$^2$/s and 3.1$\times$10$^{-13}$ erg/cm$^2$/s during quiescent and Event 1 in the  
covered band of the UVM2 filter. Combining this flux with a distance of $\approx$15 pc, we obtain a luminosity of  
3.2$\times$10$^{27}$ erg/s and 8.6$\times$10$^{27}$ erg/s during quiescent and flare Event 1, respectively.  
Using the derived temperature and the ratio of flare luminosity to the quiescent luminosity,  
an area filling-factor of the flare of $\approx 2.3$\% can be estimated.

Furthermore, one can attempt to estimate the volume of the flaring region in X-rays making use of the emission  
measure that is defined by $EM\sim n_e^2V$, where  
$n_e$ is the electron density. Using the EM of the hottest component from Table~\ref{obs} and  
$n_e$ from Table~\ref{oviitable}, 
we computed their loop-foot point area $A=\frac{V}{L}$, where L is the loop length from Table~\ref{loop_table}. 
The resulting filling factor of each of the flaring events range between $1-3\%$, which is 
consistent with the estimate from the UV emission.

\subsection{Quasi-quiescence in the chromospheric lines}\label{quasi_quiescence} 
 
As discussed in Sect. \ref{obsvssim}, throughout our observations 
we find significant variations in line shape and amplitude  
in H$\alpha$, H$\beta$, and the \ion{Ca}{ii} K lines  
on a time scale of ten minutes and less. These strong chromospheric lines 
also exhibit subcomponents in their cores. This indicates 
that they trace emission from different active regions, which are partially 
resolved in wavelength owing to the high rotational velocity of AB~Dor~A.  
The line emission thus results from the superposition of  
fluxes emitted by the different active regions on the stellar surface. The flux of the 
different active regions vary in time due to small flares and 
therefore lead to the very dynamic line behaviour. While the line 
shape is mostly affected  by this phenomenon, even the overall fluxes of 
the strong lines vary outside the flare events as can be seen in Fig. 
\ref{intensitymap}. This is comparable to the situation in X-rays, where many active  
flare stars do not exhibit a truly quiescent emission level, but some sort of flickering, 
sometimes called quasi-quiescence \citep{Robrade}. 
Moreover, 
the chromospheric emission in the strong lines is dominated by  
many active regions and not by the  narrower component emission of the flare. 
Especially in the H$\alpha$-line the flare emission only plays a minor role compared to the overall signal. 
 
\subsection{Rotational modulation} 
 
Our observations cover about 16 hours, translating into $\approx$1.3 stellar rotation periods.  
Furthermore, an analysis of rotational modulation in our data is hampered by the fact that 
a large part of it is highly influenced by flares. 
Visual inspection of the radio and X-ray light curves suggests that AB Dor A's emission is rotationally modulated \textit{outside} 
Events 1 to 3, see Fig.~\ref{light_curve}. However, closer inspection shows that the  
estimated maxima of the radio emission do not match the rotation period of 0.52~d by an hour or more. 
The same applies to the minima of the X-ray signal. 
Thus we conclude that we do not see any significant rotational modulation during our observations.  
 
 
\section{Summary and conclusions} 

The data set reported in this paper contains -- to the best of our knowledge --
the most comprehensive multi-wavelength data set on the ultra-fast rotator AB~Dor~A. These
data allow us to study temporal correlations of the star's magnetic activity
with unprecedented accuracy.  We focused on long duration flare events that were simultaneously
covered both in the optical and X-ray domains with very high spectral and temporal resolution. 
The X-ray data allow determining the coronal abundances, temperature, emission measure, 
and densities during the different states of activity. The recorded X-ray flare light curve shows a  
striking similarity to the flares observed previously on the same star described by \cite{guedel}. 

These clustered events suggest that they originate in 
the same loop or arcade as to what is observed on the Sun (see \citealt{kolomanski_2007} 
and their references); however, our simultaneously recorded high resolution optical data clearly 
show that these events must have originated in different active regions at least at the 
chromospheric level. For Events 1 and 3 we do have clear counterparts 
in the different wavelength bands except for the radio emission that is not covered and 
that exhibits no flaring activity. On the other hand, a physical connection between 
the re-heating like Event 2 in X-rays, and the coinciding event in the chromospheric lines is not clear. 
Additionally, we have found -- as in previous observations -- strong prominence-like 
absorptions in the chromospheric line profiles; however,  
we find no evidence for any X-ray absorption effects by these prominences. 
Furthermore, no signs of 
rotational modulation either in X-ray or in radio emission were found in our data.

In our high time-cadence spectral series, we could identify shallow emission subcomponents in the H$\alpha$ and 
\ion{Ca}{ii} K lines during the first few minutes of the large flare, which we 
interpreted as the signature of the strongest chromospheric shocks caused 
by the flare. Moreover, for the first time to our knowledge, we could show that during the early  
phase of a large flare both turbulent broadening for the helium and metal lines and Stark broadening for the hydrogen lines is at work, 
but later on only turbulent broadening seems to be present.
The presence of Stark broadening in the Balmer lines 
demonstrates that the flaring region exhibits high densities at the chromospheric level, as expected, 
in the context of a evaporation scenario, and the same applies to the X-ray emitting plasma, which
also shows quite high density. The filling factors derived both from the chromosphere and corona
agree and suggest a flaring region covering up to a few percent of the stellar surface, i.e., of the
order of a 1-2 $\times 10^{10}$ cm, in agreement with simple estimates derived from 
hydrodynamic loop models.
Our findings demonstrate the extreme complexity of the flare phenomenon and 
the importance of both high time-cadence and spectral resolution for stellar flare 
studies.
 
\begin{acknowledgements} 
S.~L. acknowledges funding by the DFG in the framework 
of RTG 1351. We thank Dr. Stefan Czesla for fruitful discussions.  
\end{acknowledgements} 
 
 
\bibliographystyle{aa} 
\bibliography{papers} 
\end{document}